# Map Optical Properties to Subwavelength Structures Directly via a Diffusion Model

Shijie Rao, Kaiyu Cui*, Yidong Huang*, Jiawei Yang, Yali Li, Shengjin Wang, Xue Feng, Fang Liu, and Wei Zhang

* Correspondent authors: kaiyucui@tsinghua.edu.cn, yidonghuang@tsinghua.edu.cn

Affiliation: Department of Electronic Engineering, Beijing National Research Center for Information Science and Technology (BNRist), Tsinghua University, Beijing, China

## Abstract

Subwavelength photonic structures and metamaterials provide revolutionary approaches for controlling light. The inverse design methods proposed for these subwavelength structures are vital to the development of new photonic devices. However, most of the existing inverse design methods cannot realize direct mapping from optical properties to photonic structures but instead rely on forward simulation methods to perform iterative optimization. In this work, we exploit the powerful generative abilities of artificial intelligence (AI) and propose a practical inverse design method based on latent diffusion models. Our method maps directly the optical properties to structures without the requirement of forward simulation and iterative optimization. Here, the given optical properties can work as "prompts" and guide the constructed model to correctly "draw" the required photonic structures. Simulation and experiments show that our direct mapping-based inverse design method can generate subwavelength photonic structures at high fidelity while following the given optical properties, such as transmission power, phase, and polarization responses. This may change the method used for optical design and greatly accelerate the research on new photonic devices.

## Introduction

Subwavelength structures such as photonic crystals and metamaterials have led to revolutionary approaches for light field regulation [1-4]. Because these subwavelength structures cannot be analytically modelled by geometric optics or wave optics, conventional design approaches are mainly achieved by choosing an optimal match from a predefined library with limited design space [5-8] of photonic structures based on forward simulation. The newly developed inverse design approaches [9] can greatly enlarge the design space and have shown powerful capabilities for generating less intuitive but more effective photonic structures [10, 11]. However, because existing inverse design strategies usually transform an inverse design problem into an optimization problem and obtain the optimal design by iterative algorithms [12-18], the generated photonic structure in each iteration needs to be forward modeled by numerical simulation methods such as the finite-difference time-domain (FDTD) [19], which is computationally expensive. Besides, these inverse design methods potentially face the common problems encountered by optimization algorithms involving convergence, efficiency, and global optima.

To break the limitations of the optimization-based inverse design algorithms, several attempts have been made to train deep neural networks (DNNs) and achieve a direct mapping from optical properties to photonic structures [20-34]. Recent research on artificial intelligence (AI)-generated content [35] and AI for science have shown great potential of DNNs in various practical fields, such as the generation of realistic images [36], chatbots [37], medicine [38], chemical research [39], and mechanical research [40]. To realize a practical AI for optics, we need to build a bridge between optical physics and neural network parameters. Despite the most advanced AI-based inverse design methods have demonstrated great improvements, they still encounter substantial challenges in practice such as the non-uniqueness or existence of solutions, the fabrication constraints, limited generalizability, and the different distribution of input data between training and deployment. Therefore, they are usually performed in a predefined and limited design space where the inverse problem is a one-to-one mapping [20-25]. In summary, the lack of effective and accelerated design methods has become a major concern regarding the further development of new photonic devices [1, 9, 10].

In this paper, we propose an inverse design method to achieve direct mapping from optical properties to photonic structures based on a latent diffusion model [36], named artificial intelligence-generated photonic structures (AIGPS). We exploit the powerful image synthesis ability of diffusion models [41, 42] to enlarge the search space and design new photonic structures. Here, the given optical properties can work as "text prompts" and guide the model to "draw" the required photonic structures. To achieve such a direct mapping, we also devise an encoding method of optical properties and design a prompt encoder network to solve the non-uniqueness problem and provide an interface for designing photonic structures on demand. Also, a fast forward prediction network is proposed to greatly accelerate the simulation process and realize end-to-end training. Besides, we present a training dataset containing arbitrary shapes, which empowers as large a design space as possible while meeting fabrication limits. Compared with existing methods, the diffusion network is much easier to train and more powerful than generative adversarial networks (GANs) [43]. It can be easily scaled to large sizes and provides strong generative capabilities. Moreover, our diffusion network can generate freeform shapes with various topologies. And the prompt encoder network solves the gap between abstract design demands and realistic optical properties. A more detailed comparison with existing works can be found in Supplementary Note 1.

As photonic devices are generally composed of individual subwavelength structures or arrays of building blocks which are usually referred to as meta-atoms [5, 6, 44-48], we illustrate the powerful direct mapping capabilities of our method with the example of designing meta-atoms from a given transmission power, phase, and polarization properties. Moreover, our method can be easily generalized to other photonics inverse design problems via transfer learning.

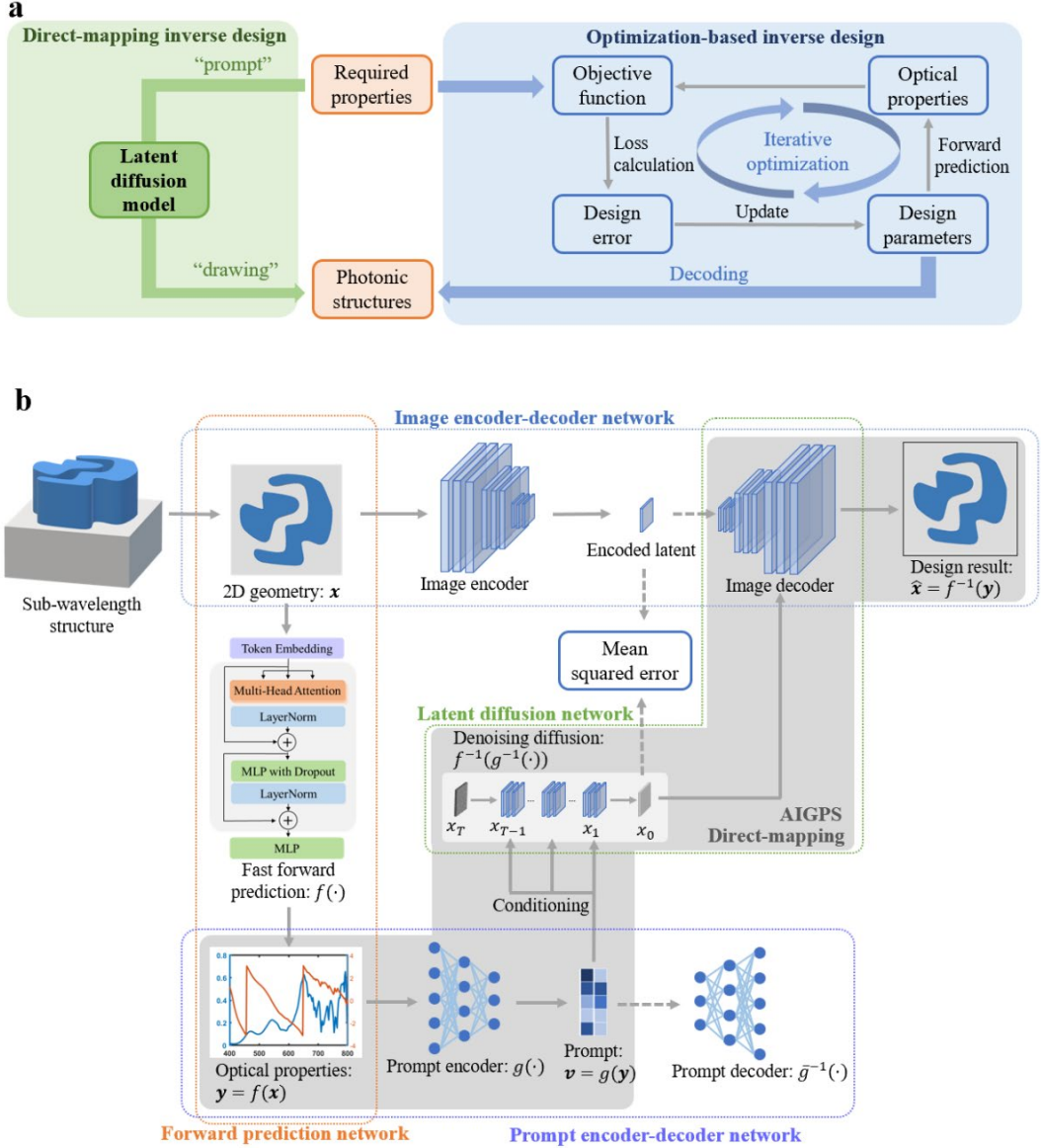

**Fig. 1 Schematics of inverse design methods. a,** Optimization-based inverse design methods usually model only the forward prediction process and execute iterative optimization algorithms to obtain an inverse design result. Instead, direct mapping-based inverse design methods aim at modeling the inverse problem of forward prediction directly. **b,** The framework of our proposed direct-mapping inverse design method is based on latent diffusion. It mainly consists of an image encoder-decoder network (MLP: multilayer perceptron), a forward prediction network, a prompt encoder-decoder network, and a latent diffusion network. Note that only the image decoder, prompt encoder, and latent diffusion are needed at inference.

## Results and Discussion
**Direct-mapping inverse design via a latent diffusion model**

Assuming that the given photonic structure is $x$ and the corresponding optical property is $y$, optimization-based inverse design methods model the forward prediction process $f(\cdot)$ such that $y = f(x)$ and perform the inverse design procedure by solving the optimization problem $\hat{x} = \underset{z}{minimize} \, ||y - f(z)||$. As shown in Fig. 1a, for optimization-based inverse design, the optimization process usually requires a large number of iterative calculations, which is computationally expensive. Instead, the direct mapping-based inverse design method attempts to model the inverse function of $f(\cdot)$ such that we can directly obtain the inverse design result $\hat{x} = f^{-1}(y)$ without any iteration. Our implementation for obtaining such an inverse function is shown in Fig. 1b. A subwavelength structure can be described by its geometry, and this geometry is usually parameterized by binary-valued pixels that indicate different materials, such as silicon and air. To accelerate the diffusion process and reduce the number of required network parameters, we adopt the latent diffusion [36] strategy, which encodes the target geometry to a set of latent parameters to greatly reduce the dimensionality of the design parameters and conduct the diffusion process in the latent space. The latent parameters generated by the diffusion network are finally decoded by the image decoder network to obtain the design results. The implementation details of our diffusion network and image encoder-decoder network are described in the Methods section.

As shown in Fig. 1b, in addition to the latent diffusion network and image encoder-decoder network, an additional forward prediction network and a prompt encoder-decoder network are also included in our inverse design framework. Although forward prediction is not needed in the direct mapping-based inverse design process, it is required during the training process to generate the input data. Therefore, to accelerate the forward prediction process, we propose a DNN-based forward prediction network to replace the numerical simulation method and efficiently train the diffusion network.

The prompt encoder network is proposed to perform fuzzy searches because we may not be able to find a photonic structure that exactly matches the given optical property, but we try to find compatible solutions in most cases. The prompt encoder network is trained to encode only the required key features of the input optical properties while ignoring other irrelevant details via self-supervised learning. The encoded results work as conditional controls for the diffusion process and guide the diffusion network to generate subwavelength structures that match the required key features. To be more specific, assuming that the prompt encoder network is $g(\cdot)$, its output $v = g(y)$ indicates the reduced properties that contain only key features such as the working band, cutoff frequency, and resonance point of optical properties. In this way, the diffusion network is not trained to model $\hat{x} = f^{-1}(y)$ but rather to model $\hat{x} = f^{-1}(g^{-1}(v))$. Notably, $v = g(y)$ is designed as a many-to-one mapping because optical properties such as the transmission responses of different meta-atoms can have the same key features. Therefore, by learning the one-to-many mapping function $g^{-1}(\cdot$

), our inverse design method can solve the non-uniqueness problem in a free rather than limited design space. Moreover, we can send only the key features instead of the whole accurate optical property to the inverse design algorithm. The prompt encoder network can accept abstract inputs, which enables fuzzy searches and provides an interface for on-demand inverse design. Our method is a data-driven approach, and the input training dataset is vital to the performance of our inverse design method because it determines the search space. Therefore, the training dataset is expected to contain 2D geometries that are as arbitrary as possible. Referring to the freeform shape generation method [49], we design an algorithm to build a dataset of arbitrary shapes that satisfies the imposed fabrication limits. These arbitrary shapes provide a much larger design space than the shapes generated from predefined parameters, thus enabling the inverse design of subwavelength structures with complex and superior functions.

To evaluate the capabilities of our direct-mapping inverse design, the principles and experiments of our inverse method are demonstrated via the design of meta-atoms, as meta-atoms are widely adopted in metasurfaces with various functions [5, 15, 44, 46, 50]. To be specific, we take meta-atoms based on a silicon-on-insulator (SOI) with a 220-nanometer-thick silicon layer as an example and train the network to produce an inverse design subjected to the given optical response. We also show that our inverse design method can be easily converted to design meta-atoms based on other materials, such as the commonly used 600-nanometer-thick $TiO_2$ layers on $SiO_2$ substrates. The detailed results of the inverse $TiO_2$ meta-atom design process can be found in Supplementary Note 2. As numerical simulations are time-consuming, by using only a small number of simulated meta-atoms as training samples, the transmission properties of every meta-atom can be quickly and accurately predicted by the designed forward prediction network. Note that although our inverse design method contains several different networks, which will be further introduced in the following sections, only the image decoder network, prompt encoder network, and latent diffusion network are needed for inference purposes. The training protocol can be found in the Methods section.

**Forward prediction network based on a transformer model**

Our forward prediction network builds on the vision transformer [51] model. As is shown in Fig. 2a, we slice the 2D geometry of the target meta-atom into $16 \times 16$ patches, and this strategy is similar to the mesh process in an FDTD simulation. Other network details are described in Supplementary Note 3. The network predicts the complex-valued transmission response of the meta-atom under horizontally polarized incident light. The network outputs the real and imaginary parts of the complex-valued transmission response, which can be further converted to the transmission power spectrum and phase spectrum.

To study the performance of our forward prediction network, we employ it in a parameter sweeping task. Parameter sweeps are commonly utilized in meta-atom design tasks to find the optimal design parameter. We sweep the period of the meta-atom shown in Fig. 2a from 300 nm to 700 nm. The transmission power spectrum and phase spectrum predicted by our network under different periods are shown in Fig. 2b. We also present the prediction results obtained with a period $p = 550nm$ separately in Fig. 2c for comparison with the FDTD simulation results. The network predictions and simulation results are almost identical. While the FDTD method may need over $10^5$ iterations to reach the simulation precision, our forward prediction network is a non-iterative and end-to-end method. It can greatly reduce the computational cost while maintaining high precision. Therefore, the proposed method is approximately 2000 times faster than the FDTD simulation strategy and can be even faster when accelerated by a GPU. Additional testing results and details can be found in Supplementary Note 4.

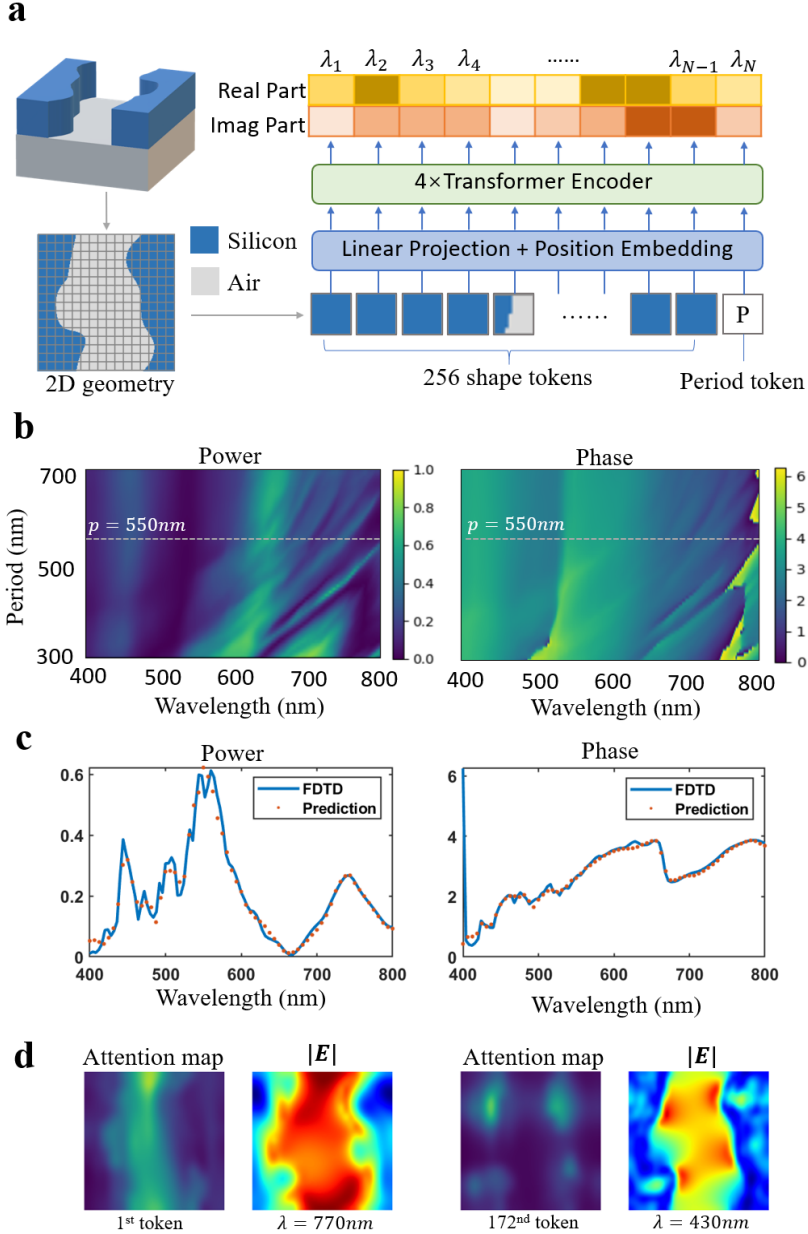

**Fig. 2 The forward prediction network and its prediction results. a.** The network is based on the vision transformer model. The 2D geometry of the meta-atom is sliced into several patches and sent to the network as inputs; this strategy is similar

to the mesh process in FDTD simulations. The period of the meta-atom is also taken as an input. **b.** The predicted transmission power and phase responses of the meta-atom are shown in **a**. We sweep the period parameter from 300 nm to 700 nm. **c.** This subfigure shows the predicted results obtained with a period $p = 550nm$, which are marked as gray dashed lines in **b**. The prediction fits well with the FDTD simulation results. **d**. The attention maps of the transformer block display some similarities to the electric field distribution at several wavelengths.

After training, the forward prediction network is fixed for training the diffusion network. Its performance indicates great potential for quickly obtaining the transmission responses of meta-atoms. Similar DNN-based fast forward prediction strategies have been adopted in previous works [17, 20, 23, 52]. However, the existing approaches usually take some predefined parameters, such as the radius and width, as inputs. In this way, the constructed DNN can only be applied to certain regular geometries. Our prediction network is designed to perform forward prediction for arbitrary shapes. It does not make any assumptions about the geometries of the subwavelength structures, thus attaining better generalizability. This method can be an alternative and efficient tool for replacing numerical simulation in studies involving metasurfaces and subwavelength structures. Furthermore, our forward prediction network has four transformer blocks, and each block can output an attention map generated by the self-attention mechanism. Analyzing these attention maps provides a possible way to understand a black-box DNN model. For example, by studying the attention maps produced by our network, we find that the first transformer block mainly extracts the interrelationships between contiguous patches, which is comprehensible because the light field is usually localized in subwavelength structures. The second block attempts to analyze the interactions between nonadjacent patches, which may help when studying nonlocal light fields. The second attention map exhibits some similarities to the electrical field distribution under certain wavelengths, as shown in Fig. 2d. The attention maps are further analyzed in Supplementary Note 5.

**Prompt encoder network trained by self-supervised learning**

As mentioned above, a prompt encoder network is vital for solving the non-uniqueness problem. We propose a self-supervised learning strategy inspired by masked autoencoders [53] to train a network that satisfies these requirements. As shown in Fig. 3a, the input of the prompt encoder network is the masking property, which is acquired by applying a certain mask to the input. In this way, some spectral bands of the transmission response are masked (where $mask = 0$) and unseen by the network. Only the unmasked (where $mask = 1$) bands are accepted by the network and encoded. Utilizing this strategy, we can mask the insignificant parts of the optical properties of interest and guide the prompt encoder network to output the reduced key features rather than the entire set of precise features.

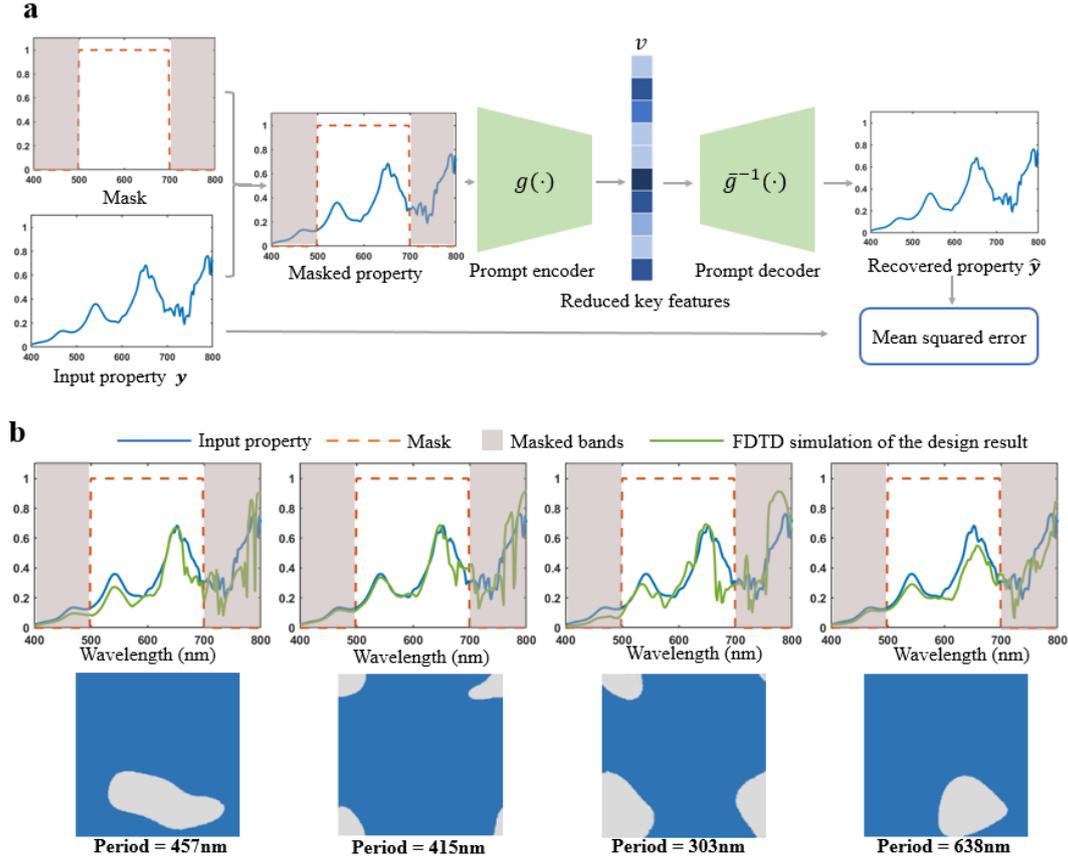

**Fig. 3 Schematic of the prompt encoder network. a.** The inputs of the prompt encoder network include the target optical property and a mask. Only the unmasked part of the property is seen by the network. An additional prompt decoder network is adopted to train the prompt encoder network via self-supervised learning. **b.** Different inverse design results that are subject to the same input property and mask are shown in **a**. The proposed prompt encoder empowers such a one-to-many mapping to solve the non-uniqueness problem.

To train a high-performance prompt encoder network, we need another decoder network that can enable self-supervised learning. The decoder network $\bar{g}^{-1}(\cdot)$ tries to recover the entire precise property according to the reduced key features $v$. Here, $\bar{g}^{-1}(\cdot)$ is designed to be a one-to-one mapping rather than a one-to-many mapping to ensure the uniqueness of its output. Self-supervised learning can be achieved by minimizing the mean squared error between the one-to-one pairs of the input property $y$ and the recovered property $\bar{y}$. The detailed training strategy, network architecture, and testing results are described in Supplementary Note 6. The training dataset includes some realistic transmission responses simulated by FDTD and some handcrafted transmission responses, including some ideal filter responses, resonance peaks, and random Fourier and Gaussian curves. In this way, we train the prompt encoder to accept both realistic data and abstract data. After conducting self-supervised learning, the decoder network is discarded, and the encoder network is fixed for training the diffusion network. Given the design target (the input property and mask) shown in Fig. 3a, after performing encoding through the prompt encoder network, the latent diffusion network can be guided to generate various design results (Fig. 3b). Each result presents a possible solution to the inverse design problem, indicating that the proposed prompt encoder enables a one-to-many mapping to solve the non-uniqueness problem.

**Inverse design results produced by the diffusion network**

After preparing the forward prediction network and the prompt encoder network, the diffusion network can be effectively trained. To test its direct mapping-based inverse design ability, we first attempt to map a full-band precise realistic transmission power response to its corresponding meta-atom, as shown in Fig. 4a. The 2D geometry and period of the meta-atom are generated directly by the diffusion network, and forward prediction is not needed. We simulate the transmission responses of the generated meta-atoms via FDTD, and they are displayed as green curves. The simulated transmission responses fit well with the design target. The results indicate great inverse design performance. Fig. 4b displays the working process of the latent diffusion network. After several denoising and diffusion steps, the required meta-atom is directly generated to not only satisfy the given transmission power properties but also meet the imposed fabrication requirements. Notably, some delicate structures that are too small to fabricate are automatically eliminated because our training dataset is designed to ensure that the generated shapes satisfy the required fabrication constraints.

C4 symmetry is an important property because meta-atoms with C4 symmetry are polarization-insensitive [40]. Our diffusion network can also be controlled to generate meta-atoms with or without C4 symmetry constraints. Fig. 4c shows the inverse design of a meta-atom with C4 symmetry. The design results also meet expectations. In addition, we can mask some uninteresting bands and generate meta-atoms that satisfy the given transmission power properties in a certain band. Some examples are shown in Fig. 4d and Fig. 4e. These results indicate that the masking strategy and the prompt encoder network provide a flexible format for the input design target.

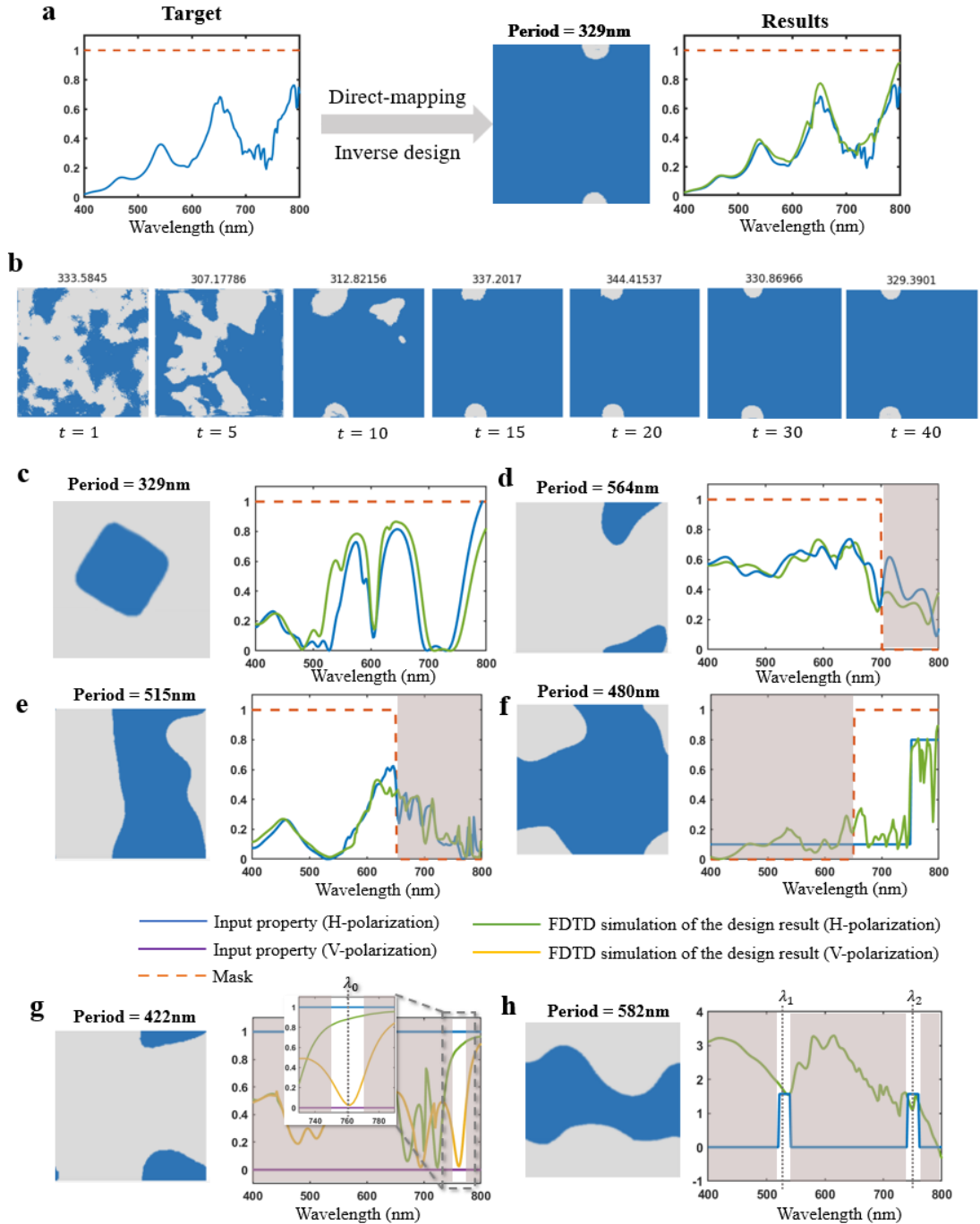

**Fig. 4 Inverse design results. a.** A full-band precise realistic transmission power response is used as the design target. **b.** The working process of the latent diffusion network. The required meta-atom is directly generated after a few denoising and diffusion steps. **c-f.** Examples of meta-atoms that are inversely designed according to the given transmission responses. **g.** Examples of inversely designed meta-atoms produced according to the transmission responses induced under different polarization conditions. **h.** An example of an inversely designed meta-atom constructed according to the transmission phase responses.

More generally, we do not fully know the precise transmission power response. Instead, we can only provide some abstract design parameters. Fig. 4f shows such an example of designing a longwave pass filter. The cutoff wavelength is expected to be approximately 750 nm, and the working band is 650~800 nm. These specifications can be easily converted to the input property (blue curve) and the input mask (orange curve) shown in Fig. 4f, which can be accepted by the prompt encoder network. Then, we can obtain the inverse design result, which is also shown in Fig. 4f. The result indicates the fuzzy search ability of our inverse design method. Such an ideal filter cannot be achieved by a 220-nanometer-thick silicon meta-atom. However, our diffusion network attempts to generate solutions that satisfy the requirements to the greatest extent possible. The performance of the meta-atom generated by our method is also verified by an FDTD simulation.

To further demonstrate the inverse design capabilities of the diffusion network, we have also employed the network to design bandstop and bandpass filters. These results can be found in Supplementary Note 8. Compared with traditional design methods such as parameter sweep, our inverse design methods can reliably generate the required meta-atoms. It can greatly accelerate and simplify the design process of subwavelength structures. For example, if we want to detect a red fluorescence protein in bioluminescent area, the best approach is to design a polarization-independent matched filter. And our diffusion network can directly map the fluorescence spectrum to meta-atoms in just a few seconds (Supplementary Note 9).

In addition to mapping the transmission power spectrum to the meta-atom, we also achieve direct mapping from the polarization properties or transmission phase spectrum to the meta-atom. Fig. 4g shows an example in which we design a meta-atom that has a high-power transmission for horizontally polarized light and a low-power transmission for vertically polarized light at a certain wavelength $\lambda_0$. Such an inverse design ability can be utilized to design a polarizer or a polarization beam splitter. Fig. 4h shows another example of designing a meta-atom that produces the same transmission responses at two different wavelengths $\lambda_1$ and $\lambda_2$; this is useful for designing phase modulation devices that work at multiple wavelengths. These results indicate the great potential of our inverse design method for use in various applications. More technical analysis regarding the latent diffusion network can be found in Supplementary Note 7.

**Fabrication and measurement results of the direct-mapping inverse-designed structures**

Fig. 4b implies that the diffusion network attempts to generate fabricable subwavelength structures, which is essential to the inverse design method. The fabrication limits are learned by the network during the training process. To test the robustness of the designed results and investigate the actual performance of our method, we designed and fabricated various subwavelength photonic structures that are shown in Fig. 5. Firstly, we mapped various transmission power responses into meta-atoms directly. We encoded a painting of sunflowers on a chip by designing 64 different structural colors. Here, the inverse design method was finetuned by transfer learning to generate meta-atoms based on silicon-on-sapphire (SOS) with a 230-nanometer-thick silicon layer. A C4-symmetry meta-atom achieves each pixel of the painting. Using our method, we mapped the desired transmission responses at three different wavelengths (623 nm for red, 528 nm for green, and 461 nm for blue) to the meta-atoms. Then, the painting can be observed directly under a microscope (Fig. 5a). Here, the color of the sapphire substrate and the fabrication error (about 10nm) result in a little chromatism between the design target and measurements.

Then, we tried to design a longwave pass filter with relatively high transmission in 700~800 nm bands and low transmission in 600~700 nm bands. The design target shown in the green dashed curve in Fig. 5b is an ideal filter, which is impossible to implement in practice. However, our method can still give an approximate solution shown in the blue curve. We fabricated the direct-mapping inverse-designed meta-atom and measured its transmission spectrum, which is shown as the red dotted curve.

Finally, we also employed the inverse design method to generate meta-atoms according to the given polarization response (Fig. 5c). We designed three different meta-atoms and encoded two different patterns into two different polarization directions at the same wavelength of 620 nm. The fabricated chip can present different patterns under horizontally and vertically polarized light as shown in Fig. 5d.

These measurement results indicate the practical capabilities of our inverse design method. It can generate the desired meta-atoms without iterative optimization. These results also show that our method can be easily generalized to design subwavelength structures of other materials by transfer learning. Moreover, during training, the diffusion model has learned the fabrication limits. Therefore, the generated structures are fabricable. More details about the design and testing of these structures can be found in Supplementary Note 10.

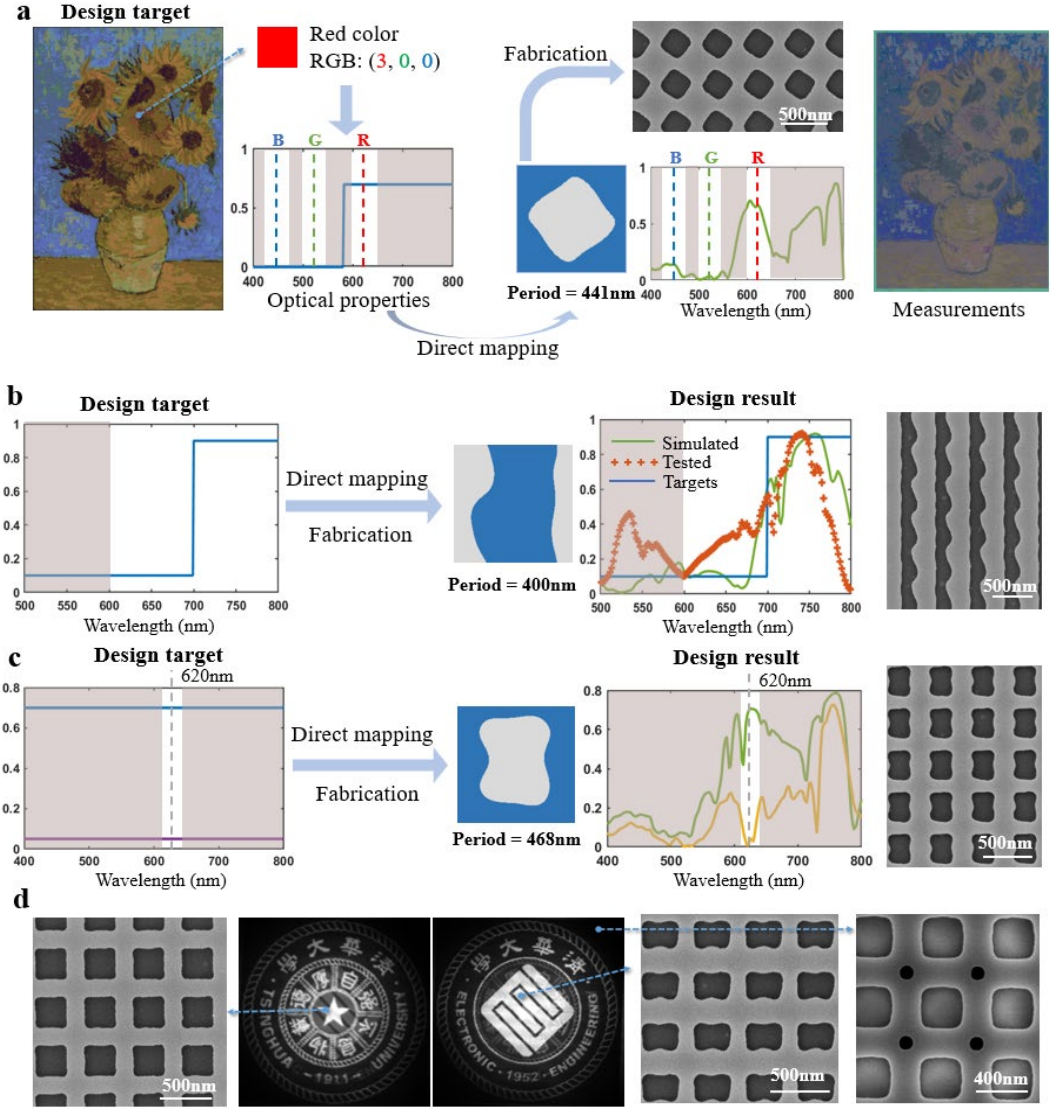

**Fig. 5 Measurement results of the inverse-designed structures. a.** A painting of sunflowers is represented by structural color. The fabricated chip is about $2.9 \times 1.9\ mm^2$ and the sunflowers can be observed directly under a microscope. **b.** An inverse-designed grating that has a high transmission at around 750 nm and a relatively low transmission at around 650 nm. **c.** An inverse-designed meta-atom that has different transmission responses under different polarization directions of incident light at around 620nm. **d.** Two different patterns are encoded into two different polarization directions at the same wavelength of 620 nm by the three inverse-designed meta-atoms.

## Conclusions

We propose a direct mapping-based inverse design method named AIGPS based on a diffusion model. While most of the existing inverse design methods are based on combinations of optimization algorithms and forward prediction strategies, our inverse design achieves direct mappings from optical properties to photonic structures, indeed providing the inverse function of forward prediction. Powered by this direct mapping technique, our method realizes a fast inverse design process without forward prediction and iterative optimization requirements. Moreover, our inverse design method is more practical than other approaches because it provides an on-demand interface for accepting abstract design parameters as inputs rather than a whole precise optical property.

One of the greatest difficulties encountered when utilizing such a direct mapping-based inverse design is the uncertainty of the solution: the solution may not be unique or may not even exist. If the solution does not exist, our method attempts to design a structure that satisfies the requirements to the greatest extent possible via a fuzzy search. However, if the given property is completely impossible to realize and no close solution can be obtained, such as in the design of a 220-nanometer-thick silicon meta-atom with high transmission at approximately 400 nm, the diffusion network may generate some random outputs because the training dataset does not contain such an impossible matching; therefore, the network does not learn to map such an input. Fortunately, our forward prediction network can be utilized to quickly evaluate the produced inverse design results. Therefore, we can determine the confidence of the inverse design results without further simulation.

In this work, we focus on inversely designing meta-atoms. Although meta-atoms have wide applications in various photonic devices, they are relatively small-scale photonic structure units and are usually studied under periodic boundary conditions. Other photonic devices, such as high-performance metalens, require the design of very large-scale photonic structures to overcome the limits of periodic meta-atoms. Our method still has the potential to solve such a large-scale inverse design problem because these large-scale structures can also be described by their geometries and compressed by the image encoder network to reduce the number of required design parameters. Latent diffusion models are also efficient at generating large-scale images. Besides, our method also has the capability of designing subwavelength structures according to multiple input properties. We only need to modify the prompt encoder network. That is, change the length/dimension of the input feature vector. In this way, different required properties are concatenated to a single vector and can control the diffusion process as well.

Moreover, our method has great potential to be scaled up to large-size models. The prompt encoder and forward prediction networks are based on Transformer, and Transformer is the basic building block in large language models such as ChatGPT. The image encoder and diffusion networks are built on latent diffusion model (LDM), which also have already been scaled up by Stable Diffusion. Therefore, our work has indicated that Transformer-based and generative deep learning models have great potential in photonics research. Given sufficient training data, it is possible to achieve a large photonics model. It may inspire future AI techniques and research, further empowering AI for photonics.

# Methods

**Implementation of the Latent Diffusion Network**

The overall architecture of our denoising diffusion model builds on the LDM [36]. We make some modifications and improvements so that the LDM is suitable for generating meta-atoms with limited training data, and we adopt the TensorFlow [54] framework to implement all of the DNN models. Here, we briefly introduce our denoising diffusion model, and more technical details can be found in the 'Code availability' section. First. The 2D geometry of the target meta-atom is described by a binary image with a size of $256 \times 256$. The period of the meta-atom is normalized to the range of $(0, 1]$ and multiplied by the binary image. The image encoder-decoder network is based on a CNN with residual connections and a self-attention mechanism. The image encoder/decoder can encode/decode the image to/from the latent space with a size of $32 \times 32 \times 2$. Denoising diffusion is conducted in the latent space to greatly reduce the incurred computational cost. The diffusion network is based on the U-Net architecture with cross-attention to introduce conditional control. The sampling strategy used in our diffusion process is DDIM [42], and we adopt a continuous diffusion time, which embeds the noise rate and signal rate into the diffusion network. In this way, the number of sampling steps can be dynamically changed at inference time. In our experiments, we find that 20 steps are sufficient for effectively generating the required meta-atom.

**Freeform meta-atom generation and simulation**

The proposed direct mapping algorithm is presented in contrast to the existing iterative optimization algorithms. Such a direct mapping cannot be achieved by analytical models. Therefore, we tried to fit the inverse function of the forward simulation by a data-driven method, in which a training dataset is required. A significant computational cost is required to generate this dataset. However, this computational effort is done once and for all. Once the inverse function is established by training, we can map any transmittance to meta-atoms directly without iterative optimization or forward simulation. The training dataset directly determines the search space of the diffusion network. It is vital for training a high-performance inverse design model and ensuring that the subwavelength structures generated by the diffusion network satisfy the imposed fabrication limits. We generate approximately 200 thousand shapes as arbitrarily as possible with and without the C4 symmetry constraint. The shapes are then randomly split into training and testing sets at a ratio of 9 to 1. The periods of the meta-atoms are between 300 nm and 700 nm. The minimum curvature radius of the shapes is restricted to 45 nm to meet the fabrication limit. If we sample the period at 10 nm intervals from 300 nm to 700 nm, we obtain 41 different values. We combine these period values with 200 thousand different shapes and obtain approximately $10^7$ different meta-atoms. It is difficult to simulate all of these meta-atoms via numerical simulation. Therefore, we only simulate approximately $10^5$ meta-atoms to train the forward prediction network.

The simulation process is performed by the FDTD method under periodic boundary conditions and horizontally polarized incident light. The mesh step is set to 10 nm and cutoff precision is $10^{-5}$. The $dt$ stability factor is set to 0.99. The transmission responses of the other meta-atoms are then quickly predicted.

**Training and inference protocols**

Each meta-atom is described by a $256 \times 256$ image. The values of the pixels are normalized to $[-1, 1]$. The image is encoded in a $32 \times 32 \times 2$ latent space by the image encoder, and the latent space is then normalized to a mean of 0 and a variance of 1. Then, we can train the image encoder-decoder network. After training, the image encoder-decoder network reaches 0.004 mean absolute error and 0.028 root mean square error. The forward prediction network takes a binary image representing the 2D geometry and the period value separately as inputs rather than encoding them to form a single image. The image encoder-decoder network and forward prediction network are first trained. Then, all of the transmission responses of the meta-atoms can be predicted quickly via forward prediction, and these responses can be used as the training set to train the prompt encoder network. Finally, we fix the trained image encoder network, forward prediction network, and prompt encoder network and train the diffusion network. At each training step, a batch of shapes is randomly chosen from the training set, and for each shape, we randomly generate a period value to form a meta-atom. Then, the transmission responses are predicted, and the meta-atoms are encoded in the latent space. The transmission responses are further encoded by the prompt encoder network and used as a conditional control mechanism for the diffusion network. Finally, we train the diffusion network to denoise the latent space with guidance provided by the input conditional control and the signal-to-noise ratio. All of the networks are trained by the Adam [55] optimizer with a weight decay rate of 0.0001, and the loss is calculated as the mean squared error. At inference time, only the image decoder network, prompt encoder network, and diffusion network are necessary. However, we can also use the forward prediction network to evaluate the generated results.

## Declarations
### List of abbreviations
AI: Artificial intelligence
GAN: Generative adversarial network
SOI: silicon-on-insulator
FDTD: Finite-difference time-domain
DNN: Deep neural network
AIGPS: Artificial intelligence-generated photonic structures
SOS: Silicon-on-sapphire
LDM: Latent diffusion model
### Ethics approval and consent to participate

Not applicable.

**Consent for publication**

Not applicable.

**Availability of data and materials**

The datasets used during the current study are available from the corresponding author on reasonable request. Our codes of the proposed AIGPS method are public available at https://github.com/rao1140427950/aigps

**Availability of data and materials**

The authors declare that they have no competing interests.

**Funding**

This work is supported by the National Key Research and Development Program of China (2023YFB2806703, 2022YFF1501600). The National Natural Science Foundation of China (Grant No. U22A6004); Beijing Frontier Science Center for Quantum Information; and Beijing Academy of Quantum Information Sciences.

**Authors' contributions**

SR was a major contributor in proposing and implementing the direct-mapping inverse design method. KC and YH supervised the study and provided assistance in meta-atom simulation and device fabrication. JW helped with building the freeform-shaped meta-atom dataset. YL and SW provided advice on neural network design and training. XF, FL, and WZ provided advice on experiments and manuscript writing. SR wrote the manuscript with the help of all other authors. All authors read and approved the final manuscript.

**Acknowledgements**

The authors would like to thank Tianjin H-Chip Technology Group Corporation and Innovation Center of Advanced Optoelectronic Chip and Institute for Electronics and Information Technology in Tianjin, Tsinghua University, for their support during electron-beam lithography (EBL) and ICP etching. Thank Yue Zou for language editing. Thank Sheng Xu and Tianhao Liu for their help in device fabrication and testing.

# Supplementary Document for Map Optical Properties to Sub-wavelength Structures Directly via a Diffusion Model

Shijie Rao, Kaiyu Cui[*], Yidong Huang[*], Jiawei Yang, Yali Li, and Shengjin Wang, Xue Feng, Fang Liu, Wei Zhang

[*] Correspondent authors: kaiyucui@tsinghua.edu.cn, yidonghuang@tsinghua.edu.cn

Affiliation: Department of Electronic Engineering, Bejing National Research Center for Information Science and Technology (BNRist), Tsinghua University, Beijing, China

**Supplementary Note 1. Comparison with existing optical inverse design neural network models**

Early works on DNN-based inverse design usually adopt fully connected networks (FCNs) [20,21,23,52]. The FCNs are easier to train, but they do not have generative abilities and can only achieve a one-to-one mapping. Therefore, they are usually performed in a predefined and limited design space where the inverse problem is a one-to-one mapping.

Generative models have become an ideal DNN-based approach to solve the problem of non-uniqueness and enlarge the design space. Generative models like generative adversarial networks (GANs) can be designed to generate outputs from the given inputs and a set of random noise. The introduced randomness makes it possible to map a single input to multiple outputs. GANs based on convolutional neural networks (CNNs) are especially popular because usually photonic structures can be described graphically. Several GANs have been proposed in recent years to achieve a direct-mapping inverse design on certain photonic devices [27-29,31,33,34]. Besides GANs, CNNs, and FCNs, Transformer models have also been adopted to perform optical inverse design [32]. It is particularly efficient in designing 1D structures. The Transformer architecture also has the potential to be scaled to large size models.

These methods have shown incredible abilities but they still face some common challenges from the direct-mapping inverse design itself and the limitations of DNNs:
1. The solution of one specific inverse design problem may be non-unique, or none at all. In most cases, we cannot find the photonic structure that exactly matches the given optical property. Instead, we should try to find compatible solutions, which results in a fuzzy search problem. However, one DNN is usually trained to fit an accurate mapping function. It may not perform well on fuzzy search.
2. The design space of the DNNs relies on the training dataset. The generated photonic structures are usually restricted to a specific distribution that depends on the training dataset. The ability of GANs to create new unseen structures is not strong enough.
3. The acceptable input optical properties are limited. During the training process, the optical properties are usually acquired directly by forward prediction and fed to the network as input. In this way, the trained DNN may only accept realistic optical properties as inputs, which is not practical enough. From the perspective of the algorithm users, for example, if we want to design a long-wave pass filter, we do not know the complete actual response of the filter, but only some design requirements such as the cut-off frequency and the pass band. However, such abstract requirements may not be acceptable to DNNs because they are usually unseen during training.

We proposed several techniques to solve these problems. Different from DNNs, CNNs, or Transformer networks, diffusion network generates images from a random noise. It is a generative network rather than a discriminative network. Therefore, it has the potential to achieve a one-to-many mapping. Compared with existing works, our method has many unique features has several advantages:

1. **This is the first work achieving optical inverse design by diffusion network**, to the best of our knowledge. The diffusion network is much easier to train and more powerful than GANs. As the latest and strongest generative network, it can break the limitations of GANs, such as the difficulties in training and scaling up. Diffusion networks and Transformer-based models have been adopted and proved their superiority in modern large-size image and video generation models. Therefore, the proposed network also has the potential to be scaled to large-size models and provide extreme generative capabilities. Moreover, the proposed method can also be easily generalized to design other subwavelength structures, such as quasicrystal, aperiodic, and supercell structures.

2. **The proposed network ensures fabrication limits and provides a much larger design space.** It can generate freeform shapes with various topologies rather than regular or predefined shapes. More importantly, conforming to the fabrication limits are very important to inverse design methods. Most of the existing GAN-based inverse design methods, such as Refs. 26, 28, 29, and 30, cannot ensure the fabrication limits. However, the generated shapes are fabricable based on our method, since the fabrication limits are learned by the network during our training process. The fabrication of inverse-designed meta-atoms is also verified by experiments shown in Fig. 5.

3. **The proposed method solves the domain gap between complete optical properties and abstract design requirements by the specially designed prompt encoder network and self-supervised learning.** It is user-friendly and can accept abstract design requirements for realistic problems. The domain gap between training data and practical usage is a huge challenge for the practical usage of the deep learning-based inverse design method. Among previous works, Ref. 29 has attempted to solve this problem by proposing the 'contrast vector'. However, the 'contrast vector' is also a predefined property that has great limitations. Moreover, two networks are trained to accept normal transmission response and 'contrast vector' separately in Ref. 29. Instead, the proposed prompt encoder network in this work solves the gap between abstract design demands and realistic optical properties for the first time by self-supervised learning. We have achieved a single network that can accept both complete optical properties and abstract design requirements, as shown in Figs. 4c-f. Moreover, the neural network model can still function normally when encountering unseen inputs due to the fuzzy search ability. These features make our method truly practical.

4. **Our method is not just aimed at a specific application** (such as structural color or polarizer), but can be used as a general approach across a range of applications with the proposed mask mechanism, as demonstrated in Figs. 4c-h. It provides very high flexibility and thus empowers AI for photonics.

**Supplementary Note 2. Fine-tuning the model via transfer learning.**

In this work, our results are mainly based on the inverse design of 220-nanometer-thick silicon meta-atoms. Here, we take the inverse design of 600-nanometer-thick $TiO_2$ meta-atoms as an example to demonstrate that our method can be easily converted to design meta-atoms based on other materials and other thicknesses via transfer learning. The substrate of the $TiO_2$ meta-atoms is also $SiO_2$.

Our inverse design method involves four DNN models: an image encoder-decoder network, a forward prediction network, a prompt encoder network, and a latent diffusion network. Note that the image encoder-decoder network and prompt encoder network are relatively universal. We do not need to retrain them if the material or the thickness of the meta-atom changes. Only the training process of the forward prediction model requires the data acquired from a numerical simulation. During the training procedure of the latent diffusion model, the optical properties are quickly predicted by the forward prediction model, and only a dataset of 2D geometries is needed. Therefore, to convert our method to design $TiO_2$ meta-atoms, the most important step is to fine-tune the forward prediction model.

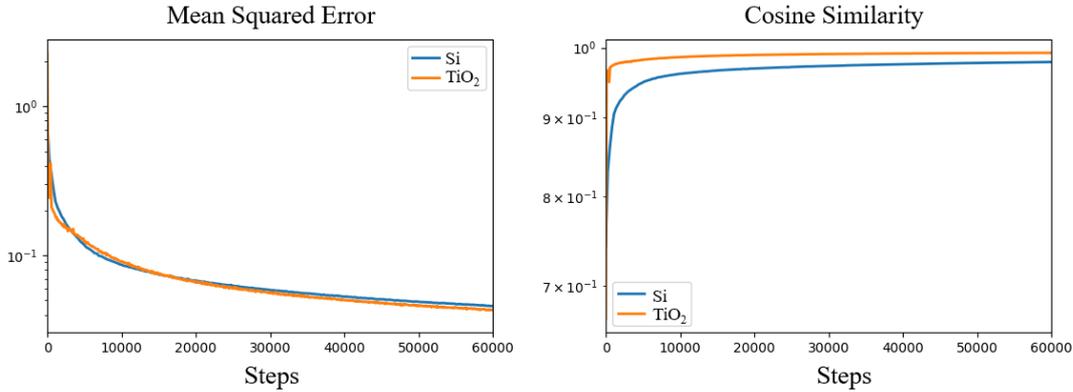

**Supplementary Fig. 1 The mean squared error and cosine similarity produced during testing.**

During the training process of the forward prediction network for silicon meta-atoms, we build a training set containing $10^5$ samples and train the network for approximately 90000 steps with a batch size of 512. To fine-tune the forward prediction network for $TiO_2$ meta-atoms, 20000 samples are simulated for constructing the training set. Then, we fine-tune the network for 60000 steps with the same batch size. The mean squared error and cosine similarity produced on the testing set during training are shown in Supplementary Fig. 1. Through transfer learning, the forward prediction network can reach the same performance as that attained before in fewer training steps while using less training data. After the training process, the mean squared error and cosine similarity reached 0.043 and 99.22% during testing, respectively.

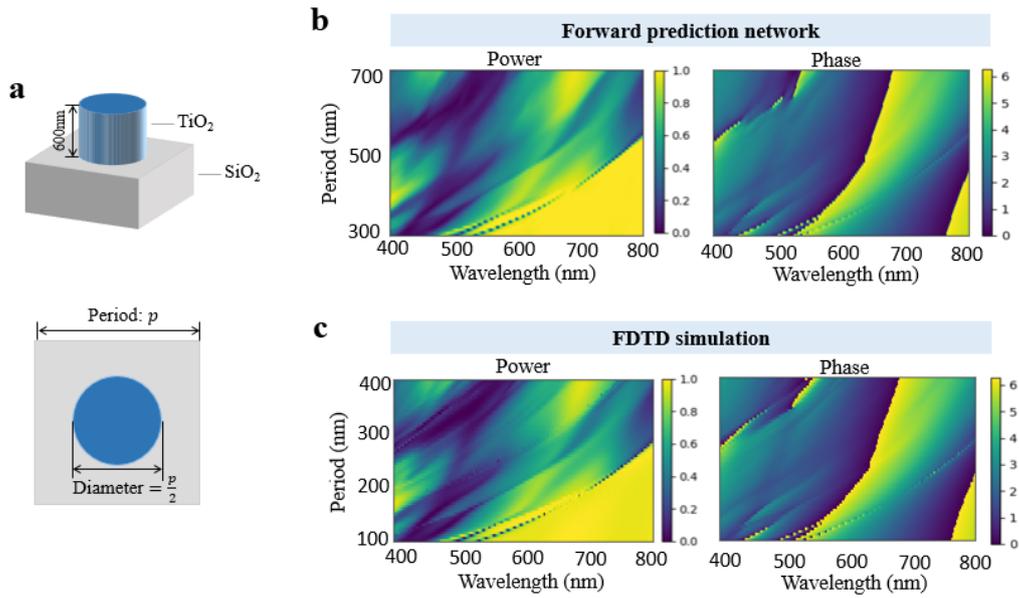

**Supplementary Fig. 2 Parameter sweeping results obtained for TiO$_2$ meta-atoms by the fine-tuned network. a.** The specifications of the meta-atom used for parameter sweeping. **b.** Parameter sweeping results obtained by our proposed forward prediction network. **c.** Parameter sweeping results were obtained via FDTD simulation.

To test the performance achieved by the fine-tuned forward prediction network for TiO$_2$ meta-atoms, we also conduct a parameter sweeping task. We sweep the period of the 600-nanometer-thick TiO$_2$ cylinder meta-atom from 300 nm to 700 nm, and the diameter of the TiO$_2$ cylinder is set to half of the period, as shown in Supplementary Fig. 2a. The sweeping results obtained from our forward prediction network (Supplementary Fig. 2b) also fit well with the FDTD simulation outputs (Supplementary Fig. 2c). These results indicate the excellent performance of our forward prediction network and the transfer learning strategy.

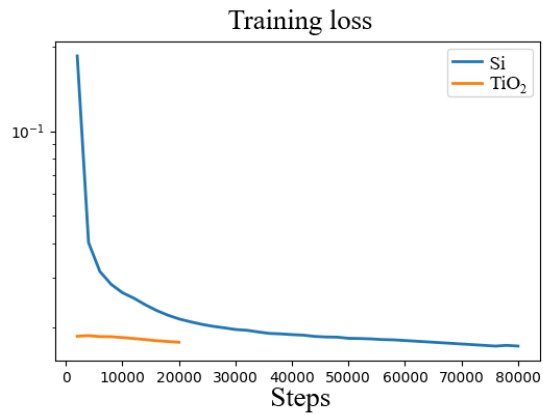

**Supplementary Fig. 3 Training losses induced by the latent diffusion network when inversely designing silicon and TiO$_2$ meta-atoms.**

The next step is to fine-tune the latent diffusion network using the fine-tuned forward prediction network. We use the same 2D geometry dataset that we employed to train the latent diffusion network for silicon meta-atoms. The training loss is plotted in Supplementary Fig. 3. We train the original latent diffusion network for silicon meta-atoms for approximately 80,000 steps with a batch size of 128 until the loss converges. During the fine-tuning process for $TiO_2$ meta-atoms, we find that the loss converges very quickly; therefore, we stop the training procedure after 20000 steps. Therefore, these experiments indicate that transfer learning can greatly reduce the incurred training cost.

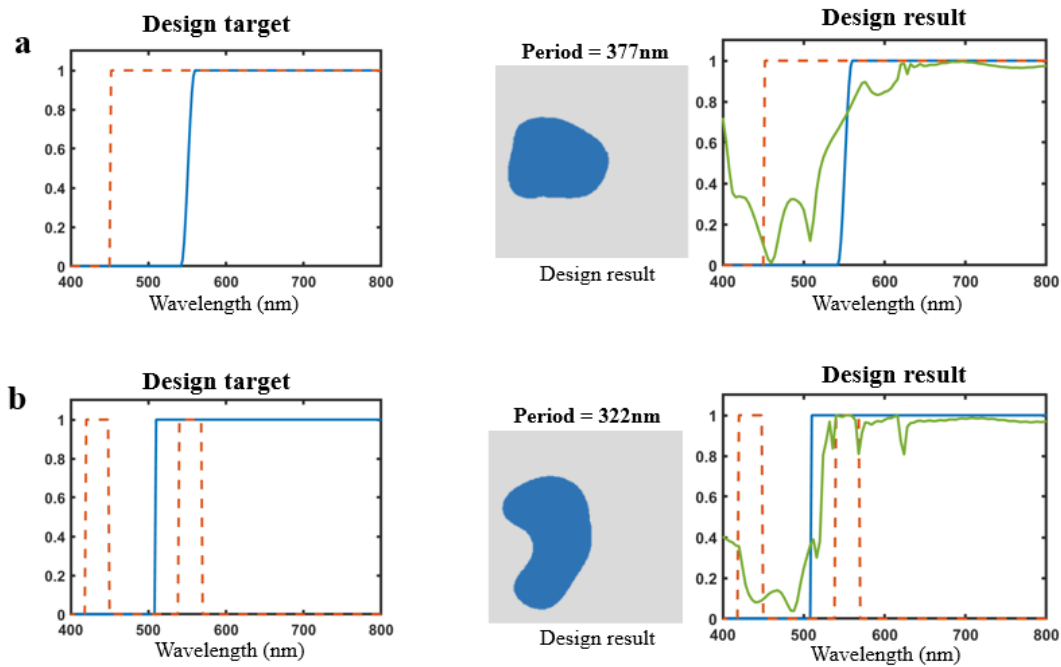

**Supplementary Fig. 4 Inverse design examples involving $TiO_2$ meta-atoms. a.** The inverse design results of a longwave pass filter. **b.** The inverse design results obtained with the requirements that the power transmission rate should be high at approximately 550 nm and low at approximately 430 nm.

To illustrate the inverse design capability of the fine-tuned network, several inverse design examples involving $TiO_2$ meta-atoms are shown in Supplementary Fig. 4. These meta-atoms are generated without the C4 symmetry constraint subject to the given transmission power responses. The FDTD simulation results (green curves) of the inversely designed meta-atoms agree well with the given design targets (blue curves). The direct mapping ability of the inverse design network trained via transfer learning can be confirmed.

# Supplementary Note 3. Design of the forward prediction network.

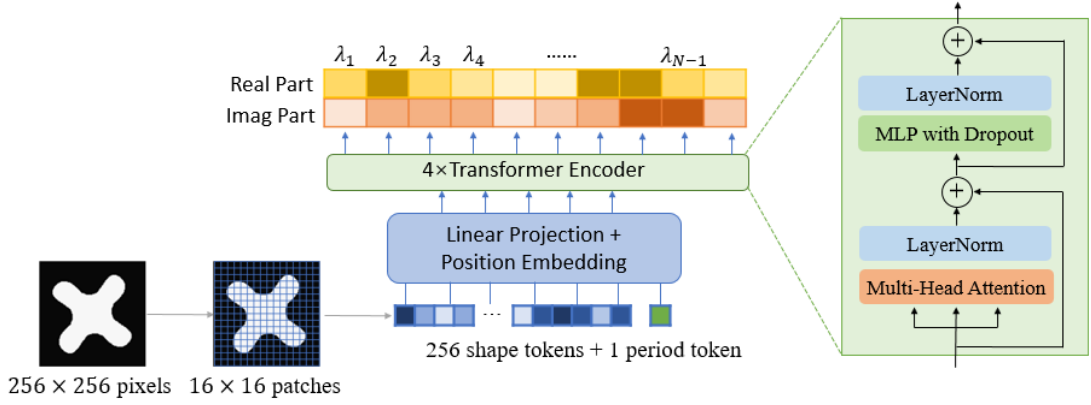

**Supplementary Fig. 5. The architecture of the transformer-based forward prediction network.**

Our forward prediction network takes the 2D geometry and period of the target meta-atom as inputs. As shown in Supplementary Fig. 5, the 2D geometry is described by a binary image with a size of $256 \times 256$. We slice the binary image into $16 \times 16$ patches. Each patch contains $16 \times 16$ pixels and is projected to a token with a feature dimensionality of 256. Each patch works as an input token of the transformer. The period of the meta-atom is randomly chosen from 300 nm to 700 nm and attached to the end of the token sequence. The period value is also projected to a token with the same dimensionality. Therefore, 256 shape tokens and 1 period token form an input feature sequence $x \in R^{257 \times 256}$. Sinusoidal positional embeddings are then added to the feature sequence. Furthermore, the feature sequence is processed by 4 transformer encoder[1] blocks. Each block has 8 heads, and the head size is set to 256. In this way, the output sequence $y \in R^{257 \times 256}$ has 257 tokens with a dimensionality of 256. Finally, each output token is mapped to a vector with 2 dimensions by another linear projection, and we can obtain the final sequence $\in R^{257 \times 2}$. The first dimension of $z$ indicates 257 sample points with wavelengths from 400 nm to 800 nm, and the second dimension contains the real and imaginary parts of the transmission response. The network is trained by the Adam[2] optimizer with a batch size of 512 for 320 epochs. The learning rate is set to 0.0002, and the weight decay rate is set to 0.0001.

It is trained to be a polarization-sensitive structure. To predict the transmission response generated under vertically polarized incident light, we can simply rotate the geometry of the meta-atom by 90 degrees and conduct prediction again using the same forward prediction network.

**Supplementary Note 4. Performance of the forward prediction network.**

According to our quantitative analysis, the mean squared error and cosine similarity of our forward prediction network reach 0.040 and 98.2%, respectively. Our forward prediction network requires approximately 42 ms on an Intel Core i7-11700 CPU @ 2.5 GHz and approximately 1.4 ms on an NVIDIA RTX 2080Ti GPU, while the FDTD simulation requires approximately 70 s on average on the same CPU. Our network, which is powered by a GPU, can operate 50000 times faster than the FDTD simulation.

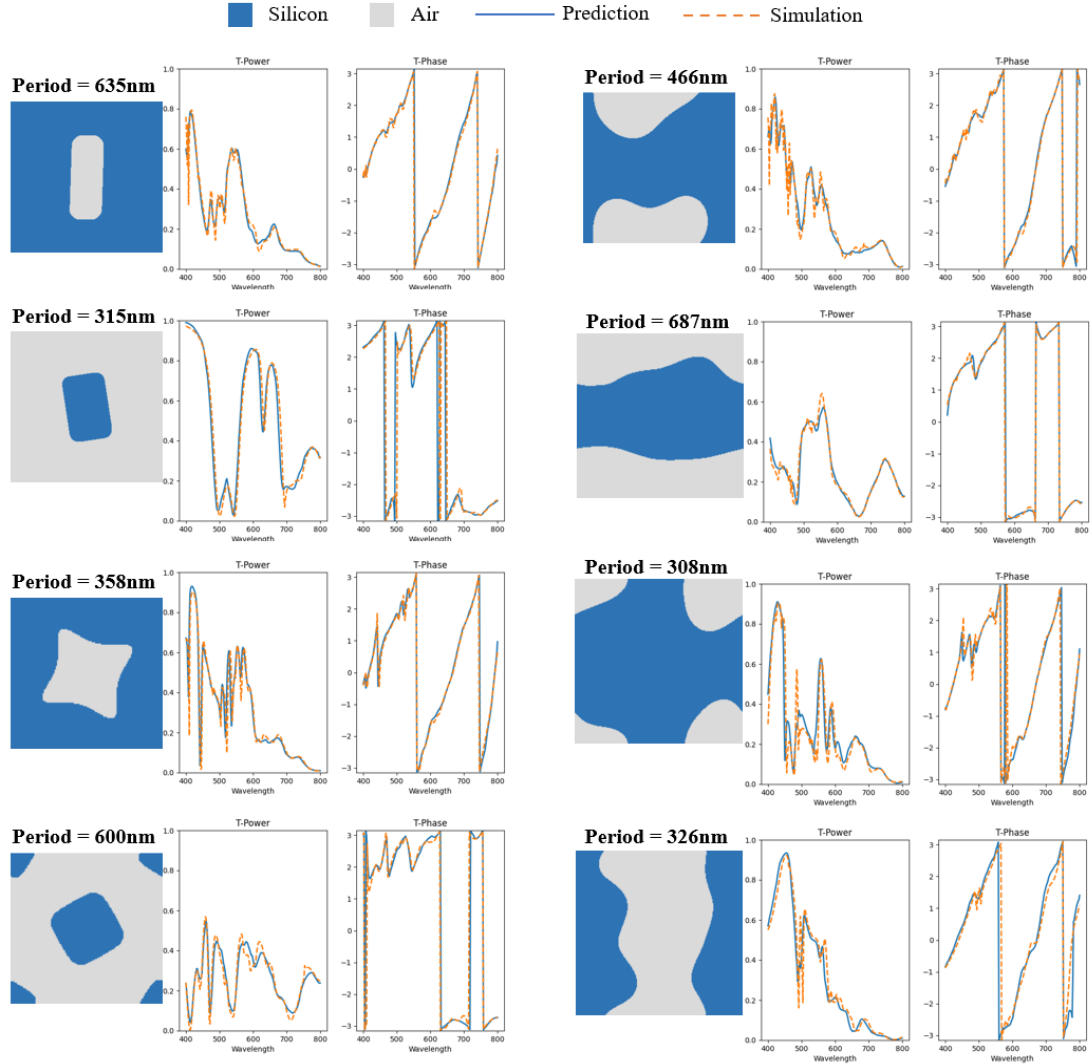

**Supplementary Fig. 6 Visualization results produced by the forward prediction network.**

A more intuitive visualization of the results is shown in Supplementary Fig. 6. We randomly select some samples from the testing set. These meta-atoms include rectangular pillars, rectangular holes, C4 symmetric pillars, C4 symmetric holes, and some randomly shaped structures. The incident light is set to horizontal polarization. The predicted transmission responses match well with the FDTD simulation outputs.

These results illustrate that our forward prediction network can greatly accelerate the forward prediction process while maintaining acceptable prediction error. This acceleration is vital for effectively training the latent diffusion network.

**Supplementary Note 5. Analysis of the attention maps produced by the forward prediction network.**

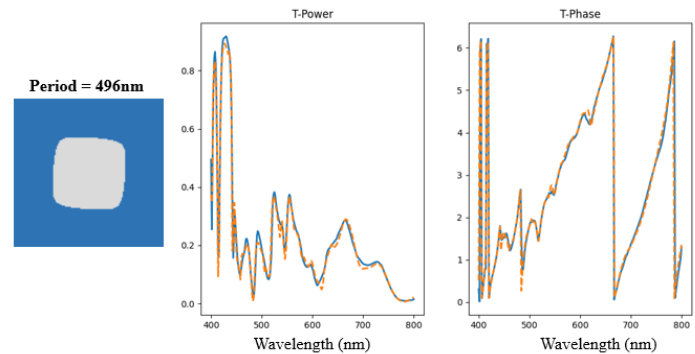

**Supplementary Fig. 7 The meta-atom used for analyzing the attention maps of the forward prediction network.**

We choose a random meta-atom as an example to analyze the attention maps yielded by the forward prediction network and try to understand what the network has learned to accurately predict the transmission response. The meta-atom is shown in Supplementary Fig. 7. The prediction results (blue curves) obtained for the transmission power and phase response highly match the FDTD simulation results (orange dashed curves).

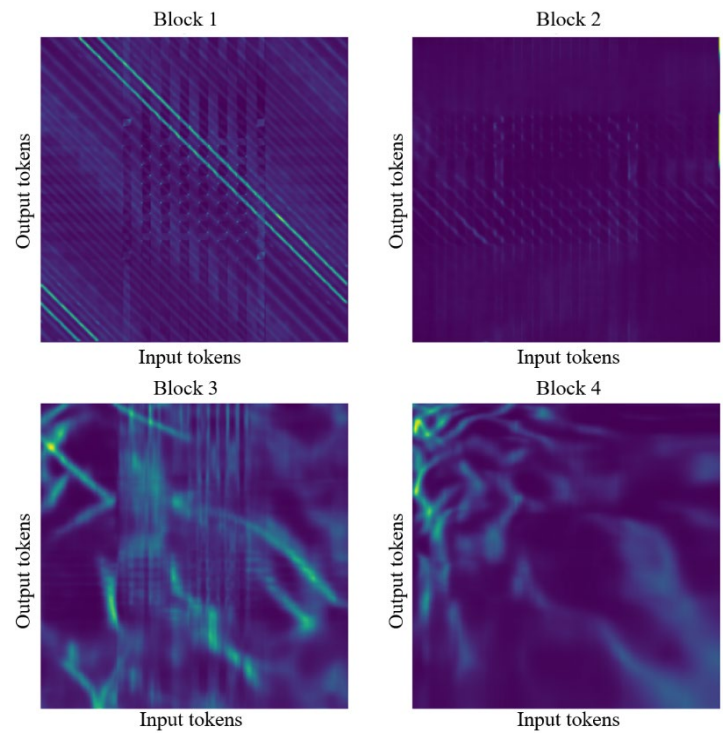

**Supplementary Fig. 8 The attention maps of the four transformer blocks.**

Four transformer encoder blocks are contained in our forward prediction network, and each block generates an attention map that maps the input tokens to the output tokens via a self-attention mechanism. The generated attention maps are shown in Supplementary Fig. 8. During the test, we find that the first two attention maps are relatively stable under different inputs, while the last two attention maps are highly related to the input data.

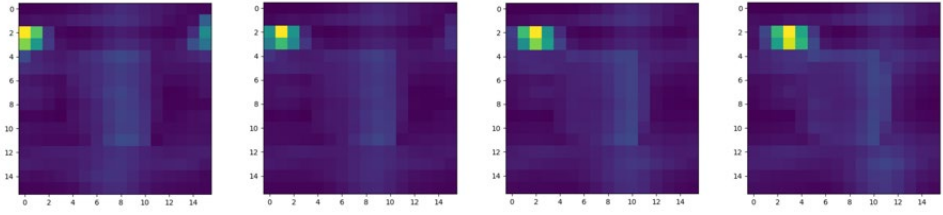

**Supplementary Fig. 9 The reshaped attention weights of the first four rows from the attention map of block 1.**

The attention map of block 1 can be effectively interpreted because the input tokens have clear physical meanings. The first 256 tokens correspond to the 256 image patches, and the last token represents the period. We can reshape the first 256 attention weights of each row of the attention map to a $16 \times 16$ matrix. Some of the results are displayed in Supplementary Fig. 9 and indicate that the first transformer encoder block mainly extracts features from the spatially adjacent patches. This is understandable because light fields are usually localized in these subwavelength dielectric structures, and the light field of each patch highly depends on its neighboring patches because these patches can provide boundary conditions. This is similar to the FDTD simulation output, where at every time step, the electrical field of each cell is calculated from the magnetic fields of the neighboring cells.

The attention map of block 2 can also be interpreted. As the output tokens of block 1, which are also the input tokens of block 2, come from the adjacent patches, we can still assume that each input token of block 2 correlates to one of the $16 \times 16$ spatial locations. Therefore, for the attention map of block 2, we can also reshape the first 256 attention weights in each row to a $16 \times 16$ matrix (Supplementary Fig. 10a). Different from the results of block 1, these results show that each output token of block 2 no longer only depends on the adjacent patches but is calculated from a relatively global feature. Each output token represents a different global feature. This can be attributed to block 2 mainly analyzing the nonlocal light field. We find that the visualized attention weight matrix (Supplementary Fig. 10a) exhibits some similarities to the electrical field distribution produced at different wavelengths for the meta-atom simulated by FDTD (Supplementary Fig. 10b). Therefore, we can assume that block 2 tries to predict the global light field distribution at each wavelength, which is important to the transmission response.

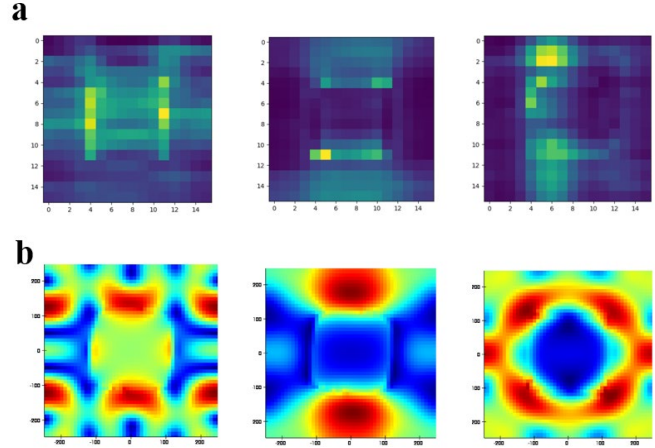

**Supplementary Fig. 10. a.** The reshaped attention weights of some rows from the attention map of block 2. **b.** The distributions of the electrical fields inside the meta-atoms simulated by FDTD at different wavelengths.

The attention maps of blocks 3 and 4 are less interpretable because we cannot find a clear physical explanation for each token. The output tokens of block 2, which are also the input tokens of block 3, no longer have a one-to-one correspondence with one of the $16 \times 16$ spatial locations. Instead, each token represents a high-level abstract feature, and finally, the tokens are transformed from the spatial domain to the spectral domain because each output token of block 4 is bound to a certain wavelength. This domain transfer task is mainly accomplished by block 4. The role of block 3 may be to map the local and nonlocal light fields to the desired optical properties.

**Supplementary Note 6. Implementation details of the prompt encoder network.**

Our prompt encoder network is implemented by the same transformer encoder block shown in Supplementary Fig. 5. The encoder network has 4 transformer encoder blocks with 4 heads, and the head size is set to 128. An additional decoder network is also designed to train the encoder network. The decoder network has 2 transformer encoder blocks with 4 heads, and the head size is 64. The inputs of the prompt encoder network are the property vector (the transmission power spectrum, transmission phase spectrum, or complex-valued transmission spectrum) and the property mask. Every element of the property vector is regarded as an input token of the transformer encoder. All of the tokens in the masked bands (where $mask = 0$) are replaced by the same trainable masked token. Sinusoidal positional embeddings are also added to the input tokens.

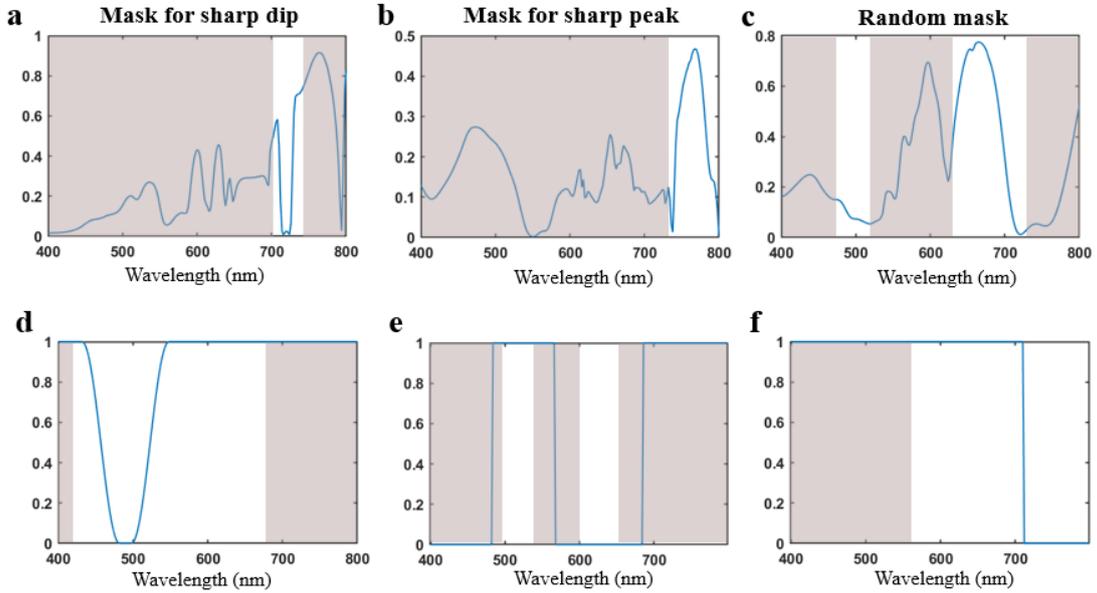

**Supplementary Fig. 11. Samples of training data for prompt encoder-decoder network. a, b, c.** Samples of realistic data. **d, e, f.** Samples of human-crafted data.

The training dataset mainly contains two types of data: realistic data and human-crafted data. The realistic data are from FDTD simulation. As for masking strategy, we apply both human-labeled masks and random masks. For human-labeled masks, we use the mask to indicate the important features in the optical properties, such as sharp dips and peaks in the transmission power responses shown in Supplementary Fig. 11a and Supplementary Fig. 11b. For random masks, we randomly generate masks at 1~3 random bands with random width, such as the data sample shown in Supplementary Fig. 11c. The human-crafted data are generated from abstract design demands. For example, the working band and cutoff frequency of a filter. We translate these abstract design demands to optical properties and apply masks that fit the demands (Supplementary Fig. 11d-f).

During training, masks are randomly generated and dropped with a dropout rate of 0.3. If a mask is dropped, it is set to all ones, and none of the input tokens are replaced by the masked token. The training data are acquired from the simulated meta-atom dataset used to train the forward prediction network. We train the networks with a batch size of 512 for 25000 steps. The Adam optimizer is also adopted to train the network. After training, the mean absolute error and root mean square error in the unmasked bands are about 0.0061 and 0.0075. Then the prompt decoder network is discarded, and the prompt encoder network is utilized to train the latent diffusion network.

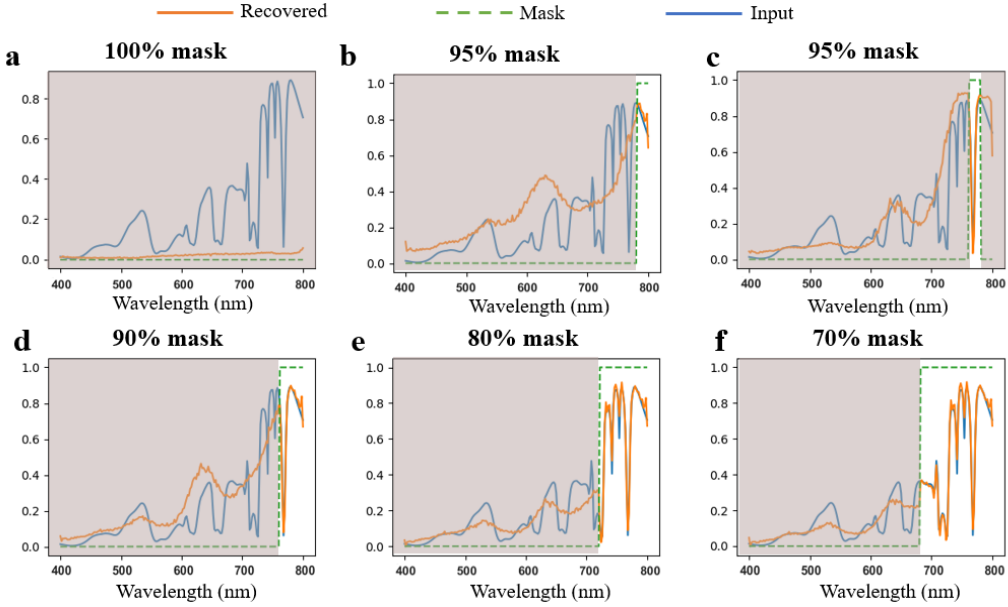

**Supplementary Fig. 12 Results of the prompt encoder-decoder network.**

While the embeddings are hard to interpret, we can visualize the decoded results to understand the behavior of the prompt encoder network. If we mask the entire bands, the outputs of the decoder are close to zeros in the entire band (Supplementary Fig. 12a). However, the values at longer wavelengths are slightly higher than those at shorter wavelengths. This is likely because, overall, the transmittance of Si-based meta-atoms is lower at shorter wavelengths and higher at longer wavelengths, and the prompt encoder-decoder network have learned this bias. If we mask 95% of the bands, the decoder can precisely recover the unmasked bands, and roughly predict the masked bands (Supplementary Fig. 12b, c). If some important features (such as digs or peaks) are unmasked, the prediction can be more accurate in the masked bands. And, if we decrease the proportion of the masked bands, the prediction can also be more accurate in the masked bands (Supplementary Fig. 12d-f).

Therefore, the proportion and position of the mask determine the possible solution space. Provided more unmasked features or more distinctive features, the solution can be found more accurately. The prediction in the masked bands may represent the average of all possible solutions. Given different mask values $\{m_1, m_2, \ldots, m_n\}$, every optical response $y$ can be encoded to different embeddings $\{g(y, m_1), g(y, m_2), \ldots, g(y, m_n)\}$ by the prompt encoder network $g(\cdot)$. If two responses $y_1$ and $y_2$ have some same features, then in the embedding space, they can have two close embeddings: $g(y_1, m_1) \approx g(y_2, m_2)$. In this way, our inverse design method can achieve fuzzy search and accept flexible inputs.

**Supplementary Note 7. Additional inverse design results provided by the diffusion network.**

First, to demonstrate the generative capabilities of our latent diffusion network, some results that are randomly generated without any conditional inputs are shown in Supplementary Fig. 13. The darker region represents air, the lighter region represents dielectric, and the degree of darkness represents the period of the meta-atom. Our latent diffusion network can generate various shapes with and without C4 symmetry.

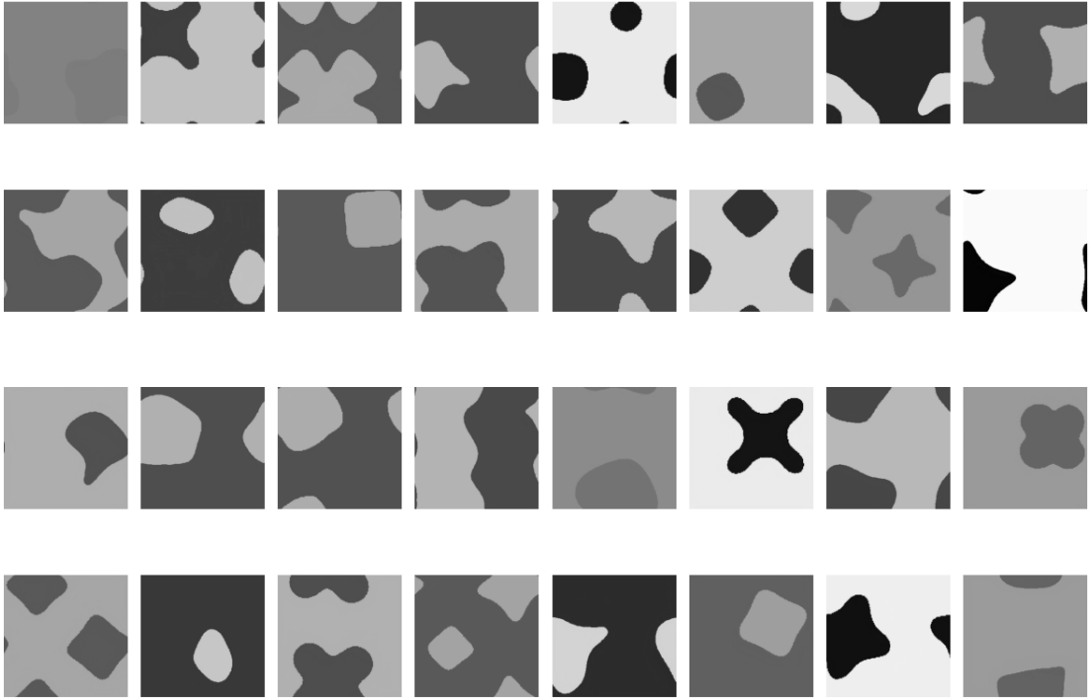

**Supplementary Fig. 13 The shapes randomly generated by the latent diffusion network.**

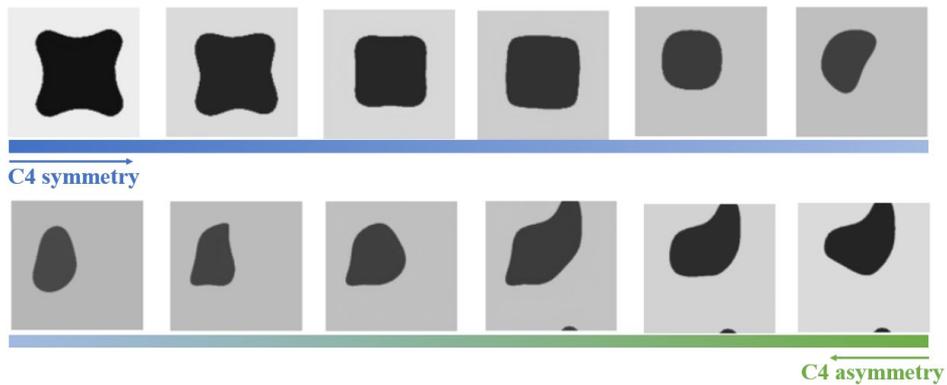

**Supplementary Fig. 14 The interpolated shapes between a C4 symmetric shape and a C4 asymmetric shape generated by a latent diffusion network.**

By proceeding through the latent space, we can see how the diffusion network utilizes interpolation to generate new shapes. We choose an initial noise value $n_1$ that can generate a C4 symmetric shape and another initial noise value $n_2$ that can generate a C4 asymmetric shape and employ the latent diffusion network to generate shapes using linear interpolations between $n_1$ and $n_2$. The interpolated shapes are shown in Supplementary Fig. 14. Diffusion models have powerful abilities to model images from different domains. The results show that our latent diffusion network can establish a method for interpolating between C4 symmetric and C4 asymmetric shapes, thus generating new shapes that are not included in the training set. This ability to generate new shapes can greatly enlarge the search space.

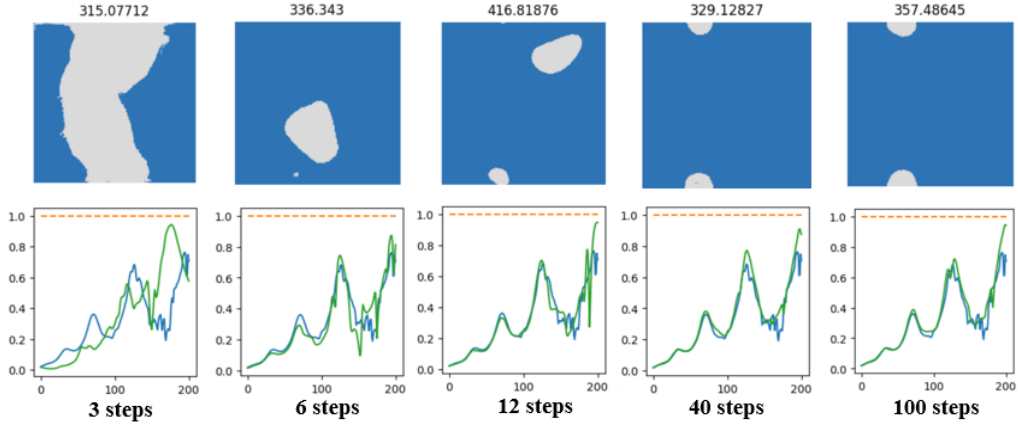

**Supplementary Fig. 15 Inverse design results obtained using different numbers of diffusion steps.**

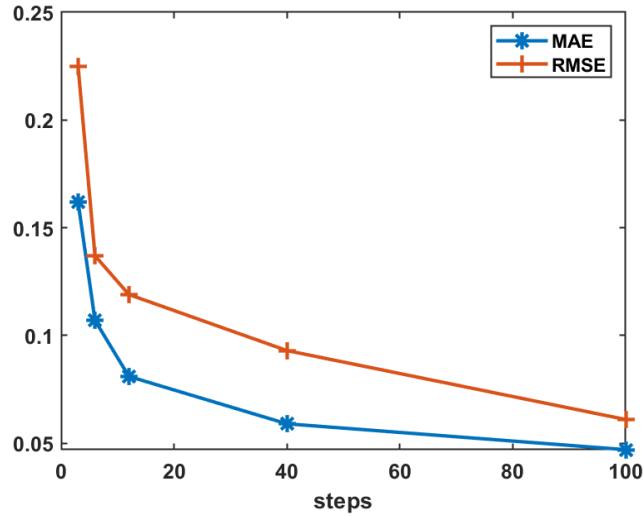

**Supplementary Fig. 16. Evaluation of inverse design performance using different diffusion steps.** We evaluate the results shown in Supplementary Fig. 15 using two metrics: MAE and RMSE. The curves indicate that the inverse design result starts to converge when $N \geq 40$. At $N = 40$, the MAE and RMSE are 0.059 and 0.093.

The number of diffusion steps is an important hyperparameter in diffusion models. As we adopt a continuous diffusion time to train the network, the number of diffusion steps $N$ can be dynamically changed during inference. A larger $N$ indicates a more refined denoising process. We conduct the same inverse design task as that shown in Fig. 4c but change $N$ to different values (the results shown in Fig. 4c are generated under $N = 40$). The inverse design results obtained under different $N$ values are shown in Supplementary Fig. 15. The first row shows the generated meta-atoms with their periods, and the second row shows the corresponding design targets (blue curves), input masks (orange curves), and design results (green curves). The results show that when $N = 6$, the inverse design result starts to satisfy the imposed requirement. When $N = 12$, we can obtain a relatively reliable inverse design result. When $N \geq 40$, the inverse design result starts to converge. Our latent diffusion network can effectively obtain inverse design results with few denoising steps.

We evaluate the results shown in Supplementary Fig. 15 using two metrics: mean absolute error (MAE) and root mean square error (RMSE). The results are shown in Supplementary Fig. 16. The curves also indicate that the inverse design result starts to converge when $N \geq 40$. At $N = 40$, the MAE and RMSE are 0.059 and 0.093. The inverse design performance of several existing methods is reported in Ref. 34. Due to the differences in the evaluation data, working bands, sampling interval, etc., the values of MAE and RMSE are not comparable directly for different methods. However, the absolute values of these metrics are similar to existing methods reported in Ref. 34, indicating that our method can achieve comparable performance while providing much more complex functions such as fuzzy search and partial inputs.

Then, we set $N = 40$, and we can see how the diffusion network denoises the initial random noise to form the desired output. This denoising process is shown in Supplementary Fig. 17.

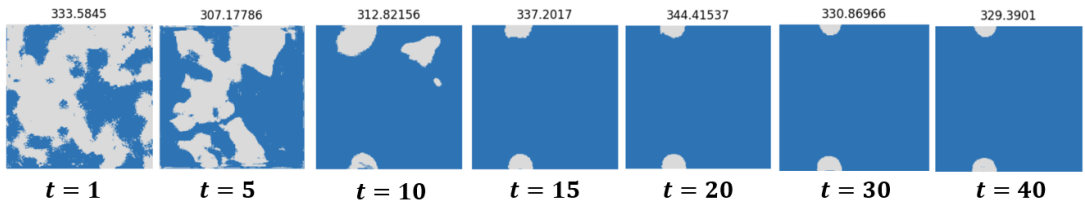

**Supplementary Fig. 17 Denoising process of the diffusion network.**

# Supplementary Note 8. Inverse design results of bandstop and bandpass filters

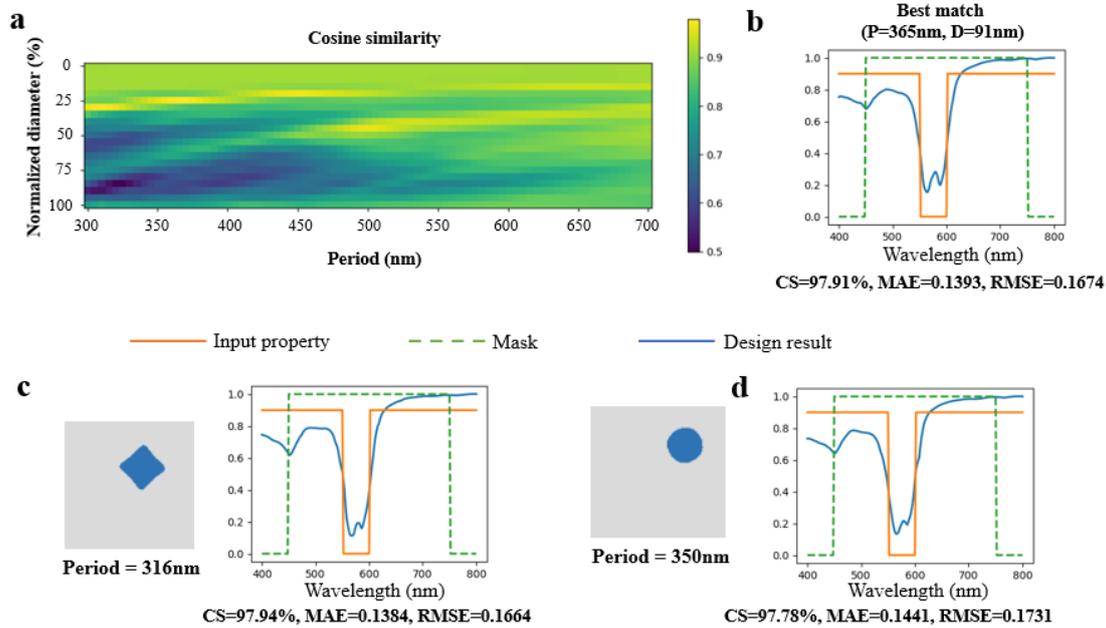

**Supplementary Fig. 18. Design results of bandstop filters by parameter sweep and diffusion network. a,** Parameter sweep results of cylindrical pillars. We sweep period $P$ from 300 nm to 700 nm at 5 nm intervals. Diameter $D$ is normalized using $P$, and we sweep the normalized diameter from 0% to 100% at 5% intervals. **b,** The best match found by parameter sweep evaluated by CS. **c, d.** The inverse design results generated by diffusion network.

If we want to design a bandstop filter (shown as orange curves in Supplementary Fig. 18) based on 220 nm SOI, a human designer may attempt to achieve the meta-atom by regular shapes such as cylindrical or rectangular pillars and find the best design parameters by parameter sweep. Taking cylindrical pillars as example, we need to determine two parameters: period ($P$) and diameter ($D$). We sweep $P$ from 300 nm to 700 nm at 5 nm intervals. $D$ is normalized using $P$, and we sweep the normalized diameter from 0% to 100% at 5% intervals. To complete the sweep, we need to run FDTD simulation 1600 times and each simulation requires several minutes. After simulation, we evaluate the difference between the design target and design results by three metrics: cosine similarity (CS), MAE, and RMSE. The sweep result of the CS metric is shown in Supplementary Fig. 18a. Then, we can find the best match by the highest CS. We find that the best match also has the lowest MAE, and lowest the RMSE, which is shown in Supplementary Fig. 18b.

The proposed direct-mapping inverse design method can make the design process much easier. We only need to set the design target to be similar to an ideal filter, and the diffusion network can output the design results in a few seconds. Although such an ideal filter cannot be physically implemented, the diffusion network can output several candidates that meet the requirements as much as possible. It can find a solution similar to the best match found by parameter sweep (Supplementary Fig. 18d). Moreover, the network can generate freeform shapes instead of only circles. Therefore, it can further achieve better results (Supplementary Fig. 18c). Note that many of the existing neural networks may fail to function properly when the input is an ideal filter. Because the response of the ideal filter does not satisfy the condition of independent and identically distributed. However, our prompt encoder network makes it possible to accept flexible inputs.

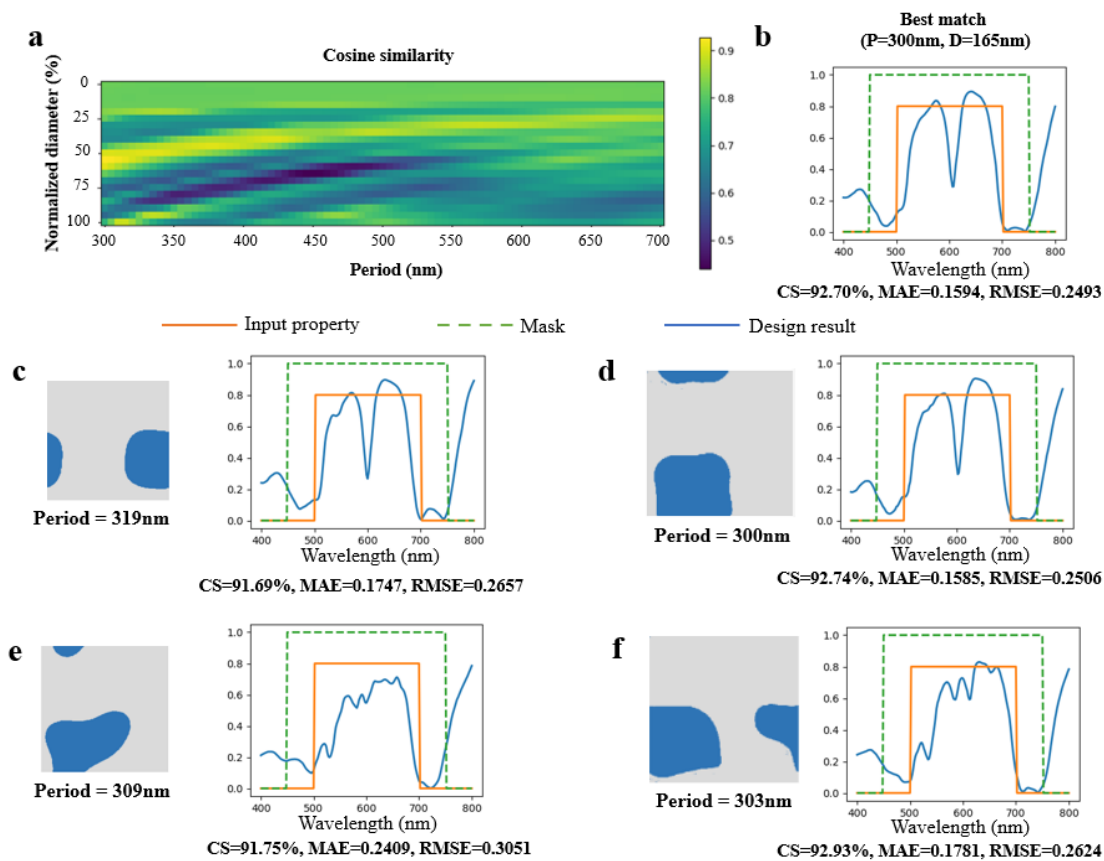

**Supplementary Fig. 19. Design results of bandpass filters by parameter sweep and diffusion network. a,** Parameter sweep results of cylindrical pillars. We sweep period $P$ from 300 nm to 700 nm at 5 nm intervals. Diameter $D$ is normalized using $P$, and we sweep the normalized diameter from 0% to 100% at 5% intervals. **b,** The best match evaluated by CS found by parameter sweep. **c-f.** The inverse design results generated by diffusion network.

To further demonstrate the inverse design capability of the diffusion network, we also employ the network to design bandpass filters (shown as orange curves in Supplementary Fig. 19). Supplementary Fig. 19b shows the result designed by the same parameter sweep method. Our diffusion network can also find a cylinder meta-atom similar to the best match found by parameter sweep (Supplementary Fig. 19c). However, these cylinder meta-atoms have a low transmittance at around 600 nm. If we remove the C4 symmetry constraint to give the diffusion network more design freedom, it can generate freeform meta-atoms directly that better meet the design target (Supplementary Fig. 19e Supplementary Fig. 19f). These results indicate that the proposed inverse design method can reliably generate the required meta-atoms. It can greatly accelerate and simplify the design process. The large design space introduced by freeform shapes leads to powerful inverse design capabilities.

## Supplementary Note 9. Inverse design results of matched filters

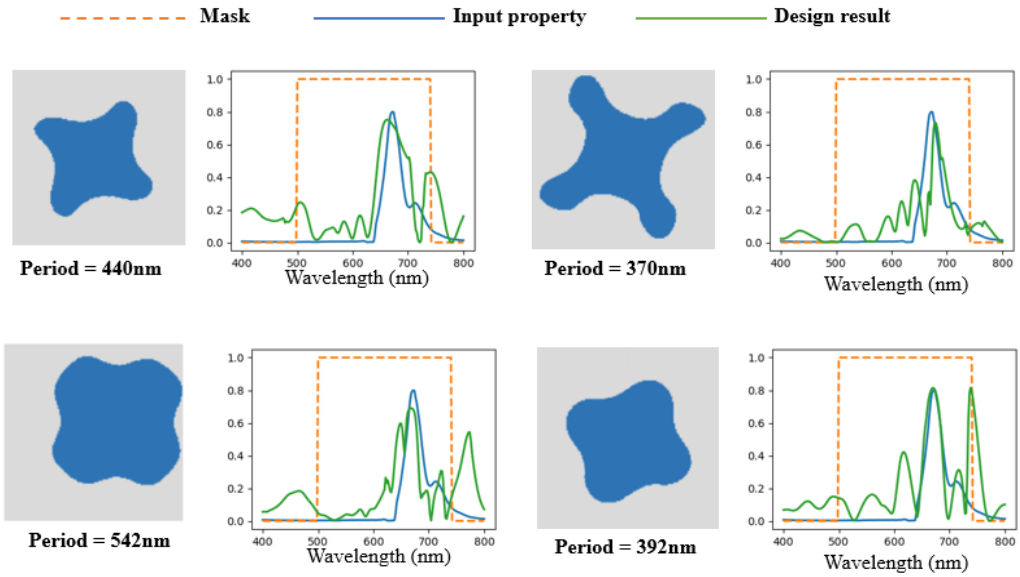

**Supplementary Fig. 20. Inverse design results of filters that can detect red fluorescence protein in bioluminescent area.**

Besides designing typical filters and polarizers, the proposed inverse design method also has advantages in other practical applications. For example, if we want to detect a red fluorescence protein in bioluminescent area (shown as blue curves in Supplementary Fig. 20), the best approach is to design a polarization-independent matched filter. While a human designer may find it difficult to design a meta-atom whose transmission response matches the fluorescence spectrum, our diffusion network can output several candidates in a few seconds (shown in Supplementary Fig. 20). It can greatly accelerate the research related to subwavelength structures.

## Supplementary Note 10. Details of fabrication and testing of the inverse-designed structures

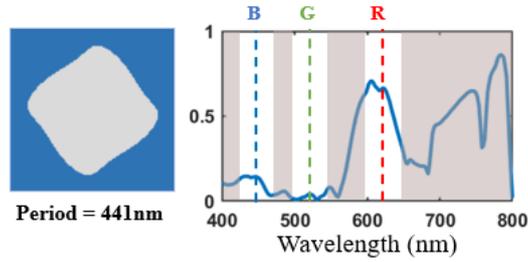

**Supplementary Fig. 21 Inverse design results of the red structural color.**

The 64 structural colors are achieved by manipulating the transmission response at three different wavelengths (623 nm for red, 528 nm for green, and 461 nm for blue). We set 4 different transmission rates at each wavelength. The rates are *(0.7, 0.46, 0.23, 0.0), (0.35, 0.23, 0.12, 0.0)*, and *(0.35, 0.23, 0.12, 0.0)* for red, green, and blue color. For example, to generate the color with a 24-bit RGB value of (255, 0, 32), we first convert the 24-bit RGB value to the 6-bit RGB value of (3, 0, 1). Then, the 6-bit RGB value is mapped to the transmission rates of *0.7, 0.0,* and *0.12* at 623 nm, 528 nm, and 461 nm. Finally, the transmission rates work as the input optical properties and guide the diffusion network to generate the desired meta-atom shown in Supplementary Fig. 21.

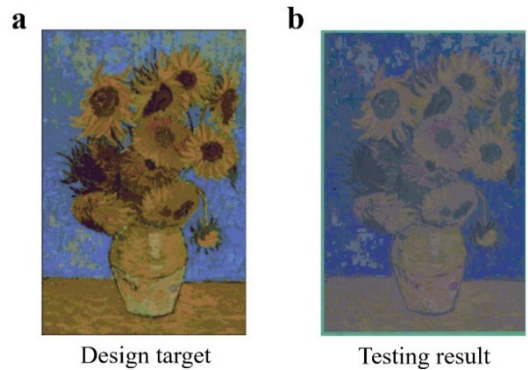

**Supplementary Fig. 22 Design target and testing result of the painting of sunflowers.**

We utilize the designed 64 meta-atoms with different structural colors to construct the painting of sunflowers shown in Supplementary Fig. 22a. The painting has $423 \times 282$ pixels. Each pixel is $6.9 \times 6.9 \ \mu m^2$ and is achieved by a specific meta-atom. After fabrication, the chip is put on an LED screen that displays the color with an RGB value of (127, 255, 255) and observed through a microscope. The microscope image is shown in Supplementary Fig. 22b. The sapphire layer results in a little chromatism between the testing result and the design target.

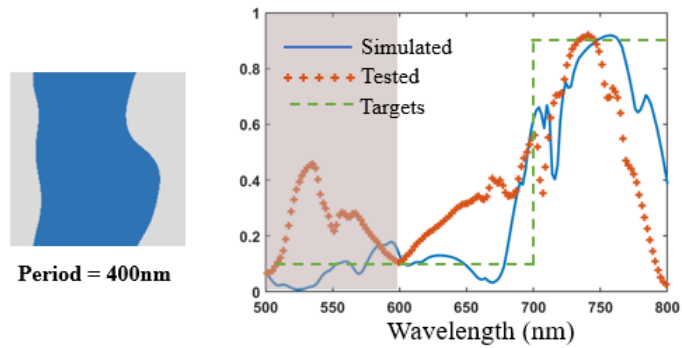

**Supplementary Fig. 23. Inverse design results of a longwave pass filter.**

The design of the longwave pass filter results in a grating-like meta-atom, which is shown in Supplementary Fig. 23. The green dashed line is the optical property condition given to the diffusion model. The model maps the optical property to a meta-atom directly, and the simulated transmission response of the meta-atom is displayed as the blue curve. We fabricated the meta-atom and measured its transmission spectrum, which is shown as the red dotted curve. The meta-atom is designed to have relatively high transmission in 700~800 nm bands and low transmission in 600~700 nm bands. The measured transmission spectrum roughly meets the design targets. However, the transmission response at 600~700 nm bands is higher than the simulation, which might be caused by fabrication error.

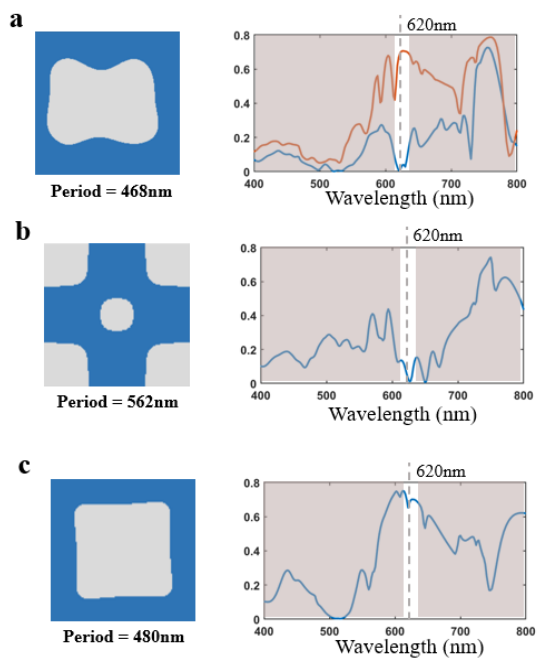

**Supplementary Fig. 24. Inverse design results of the polarization manipulating meta-atoms**

Supplementary Fig. 24 shows the inverse design results of three meta-atoms for polarization manipulation. Supplementary Fig. 24a is the designed polarization-sensitive meta-atom that has a high and low transmission for horizontally and vertically polarized light at 620 nm. Supplementary Fig. 24b and Supplementary Fig. 24c are the designed C4 symmetry meta-atoms that have high and low transmissions at 620 nm, respectively. Utilizing these three meta-atoms, we encode two different patterns into two different polarization directions at the same wavelength of 620 nm. The fabricated chip can present different patterns under horizontally and vertically polarized light.

**Supplementary Note 11. The influence of the surrogate simulator.**

We propose a forward prediction network to replace FDTD simulation. In this way, we can train the diffusion model at high-efficiency. The performance of the forward prediction network is demonstrated in Supplementary Note 5. Although it can achieve high-performance, its predictions inevitably contain some deviations. To study the influence of forward prediction network precision on the final inverse design performance, we have trained two auxiliary diffusion networks. The first one (referred to as D-fdtd) is trained using the same dataset that is adopted to train the forward prediction network. The dataset contains about $1 \times 10^5$ meta-atoms and their corresponding transmission responses simulated by FDTD. The second one (referred to as D-dnn) is also trained by the same dataset. However, we replace the simulated responses by the predictions from the forward prediction network. It is worth noting that the original inverse design network (referred to as D-0) is trained using a dataset that only contains about $2 \times 10^5$ shapes. The periods are randomly selected and transmission responses are predicted by the forward prediction network. The three networks: D-fdtd, D-dnn, and D-0, are trained using the completely same strategy.

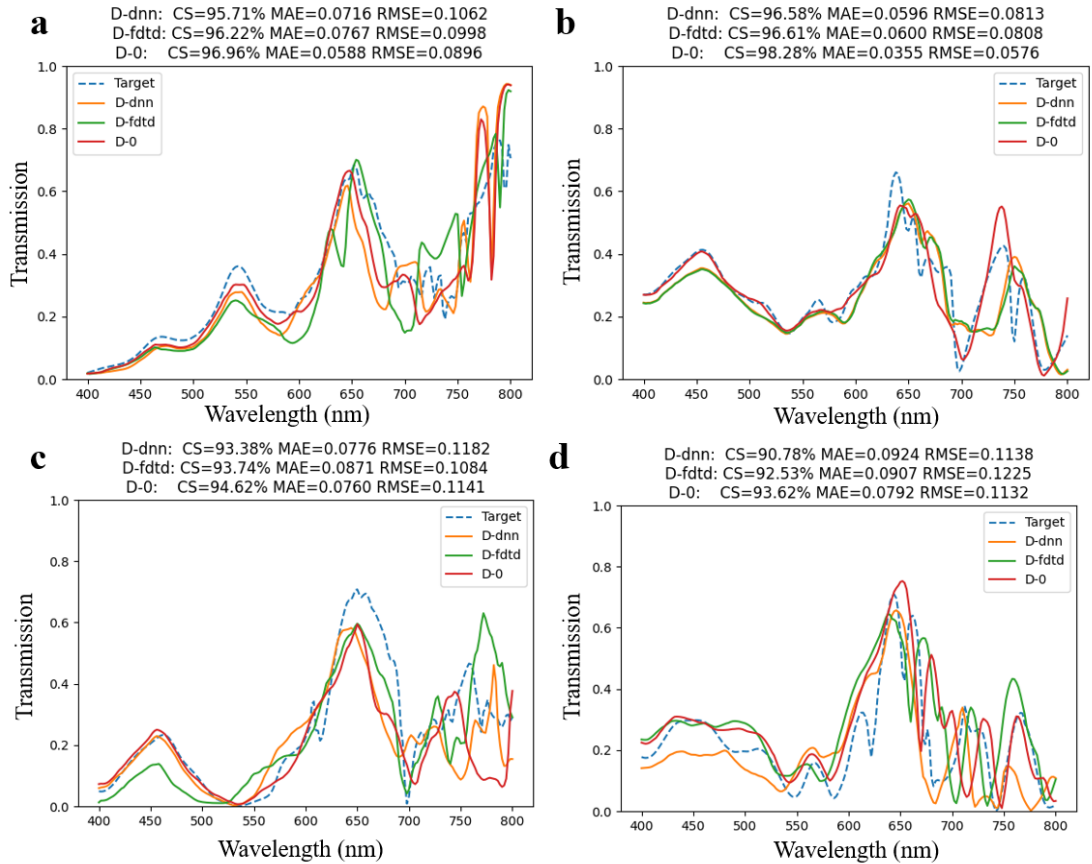

**Supplementary Fig. 25. Performance of D-dnn, D-fdtd, and D-0.**

Then, we test the inverse design performance of D-fdtd, D-dnn, and D-0. Supplementary Fig. 25 shows four inverse design results. The inverse design performance is also evaluated by CS, MAE, and RMSE metrics. The results indicate that the performance of D-fdtd and D-dnn is similar, while D-0 achieves the best results. Further, the generated shapes are encoded to latents by the image encoder network, and we evaluate the diversity of the generated shapes by the standard deviation (STD) of the latents. The results are shown in Supplementary Fig. 26a. D-0 reaches the highest STD score, indicating that its generated shapes have the highest diversity. D-0 also achieves better results on MAE and RMSE metrics. We then visualize the distribution of the generated shapes corresponding to the inverse design task shown in Supplementary Fig. 25a by PCA in Supplementary Fig. 26b, which also indicates that D-fdtd and D-dnn achieve similar diversity while D-0 has the highest diversity.

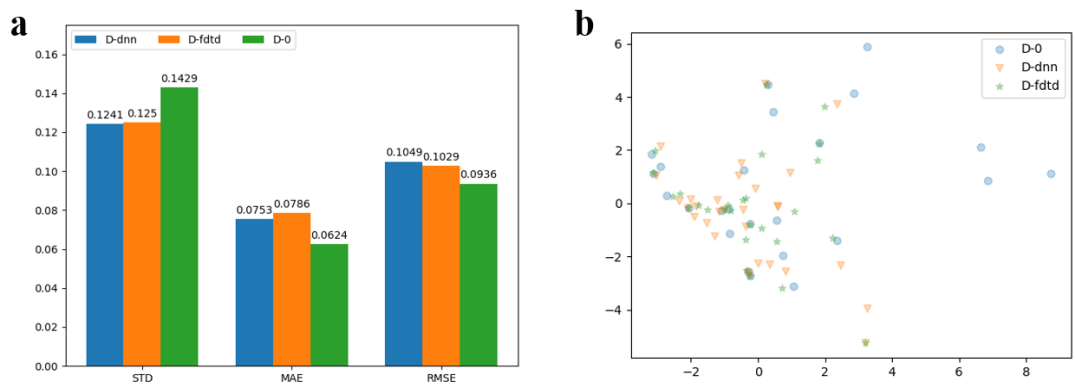

**Supplementary Fig. 26. a,** The evaluation results of D-fdtd, D-dnn, and D-0. **b,** The distribution of the generated shapes corresponding to the inverse design task shown in Supplementary Fig. 25a are visualized by PCA.

These results show that the surrogate simulator (i.e. the forward prediction network) has little impact on the final inverse design performance. Furthermore, by adopting the forward prediction network, we can utilize the whole freeform shapes dataset to train the inverse design network, thus achieving much more design freedom. And more design freedom leads to better inverse design performance and diversity as the results shown in Supplementary Fig. 26.